\documentclass[prx, notitlepage, twocolumn, nofootinbib, superscriptaddress]{revtex4-1}
\usepackage[colorlinks=true,citecolor=blue,linkcolor=blue]{hyperref}
\usepackage{graphicx}
\usepackage{color}
\usepackage{dcolumn}
\usepackage{bm}
\usepackage{mathrsfs}
\usepackage{amsmath}
\usepackage{amssymb}
\usepackage{amsthm}
\usepackage{amsfonts}
\usepackage{enumerate}
\usepackage{latexsym}
\usepackage{hyperref}
\usepackage{centernot}
\usepackage{multirow}

\begin{document}

\title{Magnetic interactions and possible structural distortion in kagome
FeGe from first-principles study and symmetry analysis}
\author{Hanjing Zhou}
\thanks{These authors contributed equally to this work.}
\affiliation{National Laboratory of Solid State Microstructures and School of Physics,
Nanjing University, Nanjing 210093, China}
\affiliation{Collaborative Innovation Center of Advanced Microstructures, Nanjing
University, Nanjing 210093, China}
\author{Songsong Yan}
\thanks{These authors contributed equally to this work.}
\affiliation{National Laboratory of Solid State Microstructures and School of Physics,
Nanjing University, Nanjing 210093, China}
\affiliation{Collaborative Innovation Center of Advanced Microstructures, Nanjing
University, Nanjing 210093, China}
\author{Dongze Fan}
\affiliation{National Laboratory of Solid State Microstructures and School of Physics,
Nanjing University, Nanjing 210093, China}
\affiliation{Collaborative Innovation Center of Advanced Microstructures, Nanjing
University, Nanjing 210093, China}
\author{Di Wang}
\thanks{Corresponding author: diwang0214@nju.edu.cn}
\affiliation{National Laboratory of Solid State Microstructures and School of Physics,
Nanjing University, Nanjing 210093, China}
\affiliation{Collaborative Innovation Center of Advanced Microstructures, Nanjing
University, Nanjing 210093, China}
\author{Xiangang Wan}
\affiliation{National Laboratory of Solid State Microstructures and School of Physics,
Nanjing University, Nanjing 210093, China}
\affiliation{Collaborative Innovation Center of Advanced Microstructures, Nanjing
University, Nanjing 210093, China}

\begin{abstract}
Recently, charge density wave (CDW) order has been discovered in a magnetic
kagome metal FeGe, providing a new platform to explore the possible
connection between magnetism and CDW in a kagome lattice. Based on density
functional theory and symmetry analysis, we present a comprehensive
investigation of electronic structure, magnetic properties and possible
structural distortion of FeGe. We estimate the magnetic parameters including
Heisenberg and Dzyaloshinskii-Moriya (DM) interactions, and find that the
ferromagnetic nearest-neighbor $J_{1}$ dominates over the others, while the
magnetic interactions between nearest kagome layers favors
antiferromagnetic. The N\'{e}el temperature $T_{N}$ and Curie-Weiss
temperature $\theta _{CW}$ are successfully reproduced, and the calculated
magnetic anisotropy energy is also in consistent with the experiment.
However, these reasonable Heisenberg interactions and magnetic anisotropy cannot explain the double cone magnetic transition, and the DM interactions, which even exist in the centrosymmetric materials, can result in this small magnetic cone angle. Unfortunately, due to the crystal symmetry of the high-temperature structure, the net contribution of DM interactions to double cone magnetic structure is absent. Based on the experimental $2\times 2\times 2$ supercell, we thus explore the subgroups of the parent phase. Group theoretical analysis reveals that there are 68 different distortions, and only four of them (space group $P622$ or $P6_{3}22$) without inversion and mirror symmetry thus can explain the low-temperature magnetic structure.
Furthermore, we
suggest that these four proposed CDW phases can be identified by using Raman
spectroscopy. Since DM interactions are very sensitive to small atomic
displacements and symmetry restrictions, we believe that symmetry analysis
is an effective method to reveal the interplay of delicate structural
distortions and complex magnetic configurations.
\end{abstract}

\date{\today }
\maketitle

\section{Introduction}

Kagome lattices are emerging as an exciting platform for the rich emergent
physics, including magnetism, charge density wave (CDW), topology, and
superconductivity \cite%
{kagome-01,kagome-04,kagome-05,kagome-03,kagome-06,mag-02,mag-03,mag-11,FeSn-03,FeSn-04,mag-01,mag-04,mag-09,mag-10,mag-05,mag-06,mag-07,mag-08,135-0,135-01,135-02,135-03,135-05,135-06,135-08,135-09,135-10,135-11,135-13,135-14,135-add1,135-add3,135-add4,135-07,135-04,135-add2,135-add7,135-add8,135-12,135-add5,135-add6,CoSn-1,CoSn-2}%
. Three key features have been identified in the electronic structure
associated with its lattice geometry, which are flat band derived from the
destructive phase interference of nearest-neighbour hopping, topological
Dirac crossing at K point in the Brillouin zone (BZ), and a pair of van Hove
singularities (vHSs) at M point \cite%
{kagome-03,kagome-06,kagome-04,kagome-05}. When large density of states from
the kagome flat bands are located near the Fermi level, strong electron
correlations can induce magnetic order \cite{kagome-04,kagome-05}. There are
several magnetic kagome materials, such as FeSn \cite%
{mag-02,mag-03,mag-11,FeSn-03,FeSn-04}, Fe$_{3}$Sn$_{2}$ \cite%
{mag-01,mag-04,mag-09,mag-10}, Mn$_{3}$Sn \cite{mag-05}, Co$_{3}$Sn$_{2}$S$%
_{2}$ \cite{mag-06} and AMn$_{6}$Sn$_{6}$ (A=Tb, Y) \cite{mag-07,mag-08},
which usually exhibit magnetic order with ferromagnetically ordered layers
that are either ferromagnetically or antiferromagnetically stacked.
Meanwhile, when vHSs are located near the Fermi level, interaction between
the saddle points and lattice instability could induce symmetry-breaking CDW
order \cite{kagome-03,kagome-06}, such as the class of recently discovered
kagome materials AV$_{3}$Sb$_{5}$ (A=K, Rb, Cs) \cite%
{135-0,135-01,135-02,135-03,135-04,135-05,135-06,135-07,135-08,135-09,135-10,135-11,135-12,135-13,135-14,135-add1,135-add2,135-add3,135-add4,135-add5,135-add6,135-add7,135-add8}%
. Significant interests have been focused on them since an unusual
competition between unconventional superconductivity and CDW order has been
found \cite%
{135-0,135-01,135-02,135-03,135-04,135-05,135-06,135-07,135-08,135-09,135-10,135-11,135-12,135-13,135-14,135-add1,135-add2,135-add3,135-add4,135-add5,135-add6,135-add7,135-add8}%
. Note that in kagome system, magnetic order and CDW order have not been
usually observed simultaneously within one material, probably due to the
fact that they originate from the flat band and the vHSs respectively, which
have the large energy difference and usually do not both appear near the
Fermi level \cite{2210.06653}.

Very recently, a CDW order has been discovered to appear deeply in a
magnetically ordered kagome metal FeGe, providing the opportunity for
understanding the interplay between CDW and magnetism in a kagome lattice
\cite%
{teng2022discovery,yin2022discovery,2203.01930,2206.12033,2210.06359,2210.06653}%
. Isostructural to FeSn \cite{mag-02,mag-03,mag-11,FeSn-03,FeSn-04} and CoSn
\cite{CoSn-1,CoSn-2}, hexagonal FeGe\ consists of stacks of Fe kagome planes
with both in-plane and inter-plane Ge atoms \cite{FeGe-1963}. A sequence of
magnetic phase transitions have been discussed in 1970-80s \cite%
{FeGe-1972,FeGe-1975,FeGe-1977,FeGe-1978,FeGe-1984,FeGe-1988}.\ Below $T_{N}$
= 410 K, FeGe exhibits collinear A-type antiferromagnetic (AFM) order with
moments aligned ferromagnetically (FM) within each plane and anti-aligned
between layers, and becomes a c-axis double cone AFM structure at a lower
temperature $T_{canting}$ = 60 K \cite{FeGe-1984,FeGe-1988}. Recent neutron
scattering, spectroscopy and transport measurements suggest a CDW in FeGe
which takes place at $T_{CDW}$ around 100K, providing the first example of a
CDW in a kagome magnet \cite{teng2022discovery,yin2022discovery}. The CDW in
FeGe enhances the AFM ordered moment and induces an emergent anomalous Hall
effect (AHE) possibly associated with a chiral flux phase similar with AV$%
_{3}$Sb$_{5}$ \cite{135-07,135-04,135-add2}, suggesting an intimate
correlation between spin, charge, and lattice degree of freedom \cite%
{teng2022discovery}. Though AHE is not usually seen in antiferromagnets in
zero field, recent studies have shown that a breaking of combined
time-reversal and lattice symmetries in the antiferromagnetic state results
in the AHE \cite{AHC-01,AHC-02,AHC-03}. In kagome FeGe, the AHE associated
with CDW order indicates that, the combined symmetry breaking occurs via the
structural distortion or magnetic structure transition below the CDW
temperature. The CDW in FeGe was then extensively studied experimentally and
theoretically \cite%
{teng2022discovery,yin2022discovery,2203.01930,2206.12033,2210.06359,2210.06653}%
, and the CDW wavevectors are identical to that of AV$_{3}$Sb$_{5}$ \cite%
{135-05,135-06,135-08,135-09,135-10,135-11}. However, sharply
different from AV$_{3}$Sb$_{5}$ \cite{135-12,135-add5,135-add6}, all the
theoretically calculated phonon frequencies in FeGe remain positive \cite%
{2206.12033,2210.06359,2210.06653}, and the structural distortion of the CDW
phase remain elusive. It is firstly\ suggested to be reduced to $P622$ with
the distortion of two non-equivalent Fe atoms \cite{teng2022discovery},
while the later works propose that FeGe shares the same space group of $%
P6/mmm$ with the pristine phase \cite{2206.12033,2210.06359}. Based on
first-principles calculations and scanning tunneling microscopy, Shao $et$ $%
al.$ show that the CDW\ phase of FeGe exhibits a generalized Kekul\'{e}
distortion \cite{kekule1865studies} in the Ge honeycomb atomic layers \cite%
{2206.12033}. Meanwhile, using hard x-ray diffraction and spectroscopy, Miao
$et$ $al.$\ report an experimental discovery of charge dimerization that
coexists with the CDW phase in FeGe \cite{2210.06359}. Therefore, the
understanding of the magnetism, and the intertwined connection between
complex magnetism and structural distortion in kagome FeGe is an emergency
issue, which we will address in this work based on first-principles study
and symmetry analysis.

In this work, we systematically analyze the electronic and magnetic
properties of kagome FeGe. Our numerical results show that this material is
a magnetic metal exhibiting large magnetic splitting around 1.8 eV. Based on
combining magnetic force theorem and linear-response approach \cite%
{J-1987,wan2006,wan2009}, the magnetic exchange parameters have been
estimated. The results show that the nearest-neighbor $J_{1}$ is FM and
dominates over the others, while the magnetic interactions between nearest
kagome layers favors AFM, consequently resulting in the A-type AFM
ground-state configuration. Based on these spin exchange parameters, the
calculated N\'{e}el temperature and Curie-Weiss temperature also agree well
with the experiments. Using the method in Ref. \cite{force-1,force-2}, we
also calculate the magnetic anisotropic energy (MAE) to be around 0.066 meV
per Fe atom with easy axis being out of the kagome layers, which is in
reasonable agreement with the experimental results \cite{FeGe-1988}.
However, the double cone magnetic transition at $%
T_{canting}$ = 60 K cannot be reproduced by these reasonable magnetic
parameters. We find that Dzyaloshinskii-Moriya (DM) interactions \cite{DM-D,DM-M}\ are much more efficient than Heisenberg interactions for causing this canted spin
structure. Unfortunately,  the space group $P6/mmm$ of high-temperature phase in FeGe has
inversion symmetry and mirror symmetries, and all of them eliminate the net
contribution of DM interactions to
the double cone magnetic structure.
It is well known that DM interactions are very sensitive to atomic
displacements, while small structural distortion usually has little effect
on Heisenberg interactions. Therefore we explore the possible CDW distortions which can explain the low-temperature magnetic structure.
Symmetry theoretical analysis reveals that there are 68 different distortions, which are the subgroups of the parent $P6/mmm$\ phase with $2\times
2\times 2$ supercell \cite%
{teng2022discovery,yin2022discovery,2206.12033,2210.06359}.
Based on the group theoretical analysis, we find that only
four structures (space groups $P622$ and $P6_{3}22$) without inversion and mirror symmetry thus can have double cone spin structure.
We further propose that using Raman spectroscopy, these four CDW phases can be identified from their different
numbers of Raman active peaks.

\begin{figure*}[!htb]
\centering\includegraphics[width=0.98\textwidth]{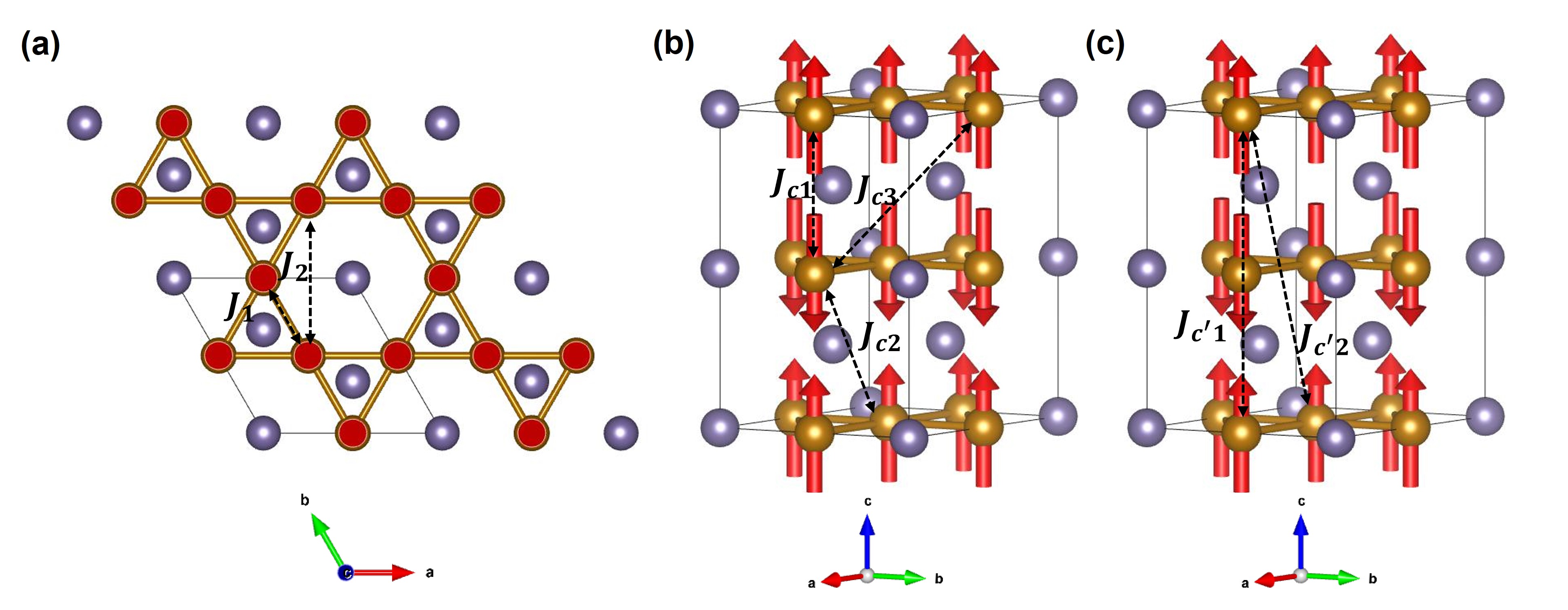}
\caption{Crystal and magnetic structures of FeGe. Yellow and purple spheres
represent Fe and Ge atoms respectively, while arrows denote magnetic moments
of Fe atoms. (a) Top view of FeGe. The exchange interactions $J_{i}\ $denote
the $i$th-nearest-neighbor interactions between Fe ions within kagome
layers. (b) The exchange interactions $J_{ci}$ denote the $i$%
th-nearest-neighbor interactions between Fe ions on the nearest kagome
layers. (c) The exchange interactions $J_{c^{\prime }i}$ denote the $i$%
th-nearest-neighbor interactions between Fe ions on the next nearest kagome
layers. }
\label{crystal}
\end{figure*}

\section{Method}

The first-principles calculations have been carried out by using the full
potential linearized augmented plane-wave method as implemented in the
Wien2k package \cite{blaha2001wien2k}. The converged k-point Monkhorst-Pack
meshes are used for the calculations depending on materials. The
self-consistent calculations are considered to be converged when the
difference in the total energy of the crystal does not exceed $0.01mRy$. We
adopt local spin-density approximation (LSDA) \cite{vosko1980accurate} as
the exchange-correlation potential, and include the spin orbit coupling
(SOC)\ using the second-order variational procedure \cite%
{koelling1977technique}.

The spin exchange interactions, including Heisenberg and DM\ interactions
\cite{DM-D,DM-M}, are calculated using first principles based on combining
magnetic force theorem and linear-response approach \cite%
{J-1987,wan2006,wan2009}, which have successfully applied to various
magnetic materials \cite{wan2006,wan2009,wan2011,wan2021,mywork-1}.

Monte Carlo (MC) simulations are performed with Metropolis algorithm for
Heisenberg model \cite{metropolis1949monte,MC-1,MC-2}. The size of the cell
in the MC simulation are 16$\times $16$\times $16-unit cells with periodic
boundary conditions. At each temperature we carry out 400000 sweeps to
prepare the system, and sample averages are accumulated over 800000 sweeps.

\section{Results}

\subsection{The electronic and magnetic properties}

\begin{figure*}[tbph]
\centering\includegraphics[width=0.98\textwidth]{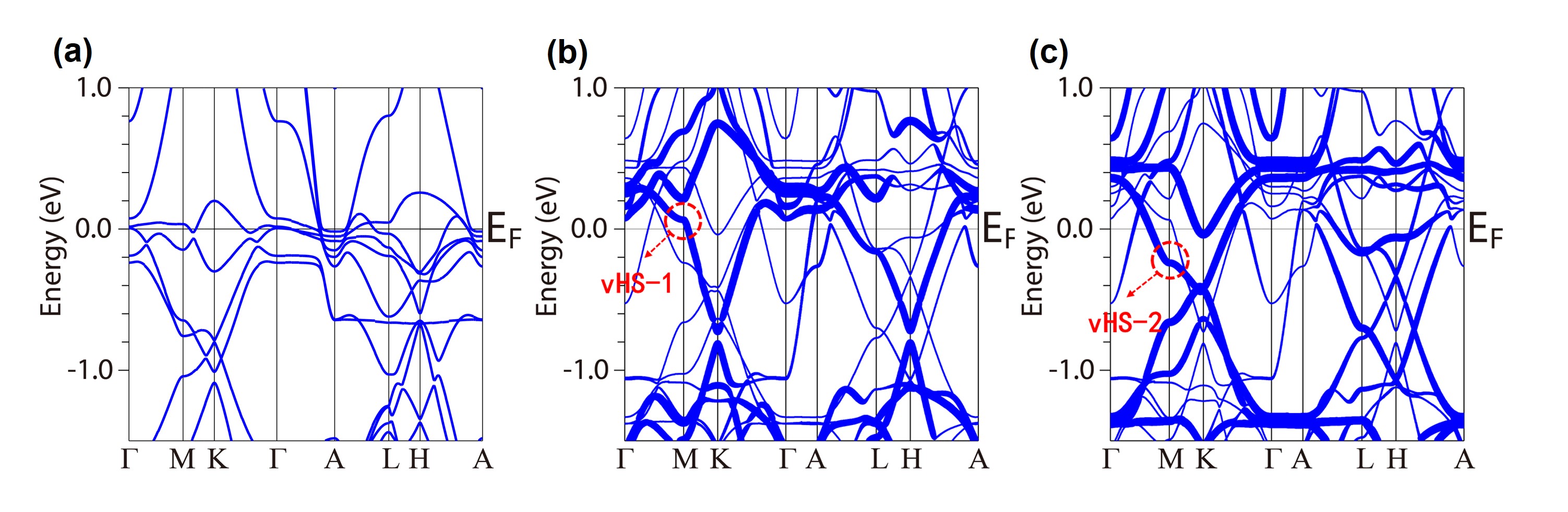}
\caption{(a) Band structure of nonmagnetic FeGe from LDA + SOC calculation.
(b),(c) Orbital-resolved band structure of Fe-$d_{xy}/d_{x^{2}-y^{2}}$ and
Fe-$d_{xz}/d_{yz}$ for A-type AFM configuration with spin orientations along
the (001) direction from LSDA + SOC calculation.}
\label{band}
\end{figure*}

The pristine phase of FeGe crystallizes in the hexagonal structure with
space group P6/mmm (No. 191) \cite{FeGe-1963}, where the coordinates of the
atoms are shown in table\ \ref{cdw} and Fig. \ref{crystal}. Firstly we
perform nonmagnetic local-density approximation (LDA) + SOC calculation, and
show the band structures in Fig. \ref{band}(a). While Ge-2$p$ states are
mainly located between -6.0 and -2.0 eV, the main contribution around the
Fermi level comes from the 3$d$ orbitals of Fe ions, as shown in Fig. \ref%
{dos} of Appendix. Consistent with previous first-principle calculations
\cite{2203.01930,2210.06653}, the kagome flat bands around the Fermi level
exhibit a large peak in the density of states, which indicates the magnetic
instability. Therefore, LSDA + SOC calculations are performed based on the
A-type AFM configuration and the band structures are shown in Figs. \ref%
{band}(b) and \ref{band}(c). The magnetic moment of Fe ions is estimated to
be 1.55 $\mu _{B}$, which is in agreement with the previous experimental
value around 1.7 $\mu _{B}$\ \cite{FeGe-1975,FeGe-1978}.\ Note that each
kagome layer is FM and the key signatures of electronic structures in kagome
lattice are remained. The magnetic splitting is around 1.8 eV (see Fig. \ref%
{dos} of Appendix), which makes that the flat bands above and below Fermi
level correspond to the spin minority bands and spin majority bands
respectively. Meanwhile, the vHSs that are relatively far from the Fermi
level in the nonmagnetic state, are brought near the Fermi level by the spin
splitting, as shown in Figs. \ref{band}(b) and \ref{band}(c). We present
orbital-resolved band structures, and find that the vHSs near the Fermi
level, which marked as vHS-1 and vHS-2 in Figs. \ref{band}(b) and \ref{band}%
(c), are mainly contributed by the $d_{xy}/d_{x^{2}-y^{2}}$ and $%
d_{xz}/d_{yz}$ orbitals respectively. These vHSs near the Fermi level are
suggested to induce symmetry-breaking CDW order in kagome metal FeGe \cite%
{2210.06653}.


\begin{table}[tbp]
\caption{Spin exchange parameters (in meV) including Heisenberg and DM
interactions of FeGe evaluated from LSDA+SOC calculations, respectively. The
Fe-Fe distances and the corresponding number of neighbors NN are presented
in the second and third columns.}%
\begin{tabular}{c|cccc}
\hline\hline
& Distance($\mathring{\mathrm{A}}$) & NN & J & DM \\ \hline
$J_{1}$ & 2.50 & 4 & -41.97 & (0, 0, 0.03) \\
$J_{2}$ & 4.33 & 4 & 5.49 & (0, 0, -0.12) \\ \hline
$J_{c1}$ & 4.05 & 2 & 8.44 & (0, 0, 0) \\
$J_{c2}$ & 4.76 & 8 & -2.04 & (0.01, -0.02, -0.07) \\
$J_{c3}$ & 5.93 & 8 & 1.81 & (0.07, -0.04, -0.09) \\ \hline
$J_{c^{\prime }1}$ & 8.11 & 2 & -0.66 & (0, 0, 0) \\
$J_{c^{\prime }2}$ & 8.49 & 8 & 0.09 & (-0.04, -0.09, -0.03) \\ \hline\hline
\end{tabular}%
\label{JDM}
\end{table}

To quantitatively understand the rich magnetic phenomenon in kagome\ FeGe, a
microscopic magnetic model with proper parameters is extremely important.
Based on the calculated electronic structures, we estimate the exchange
parameters including Heisenberg and DM interactions using the
linear-response approach \cite{J-1987,wan2006,wan2009} and summarize the
results in table \ref{JDM}. As shown in Fig. \ref{crystal}, we divide the
magnetic interactions considered into three types: the exchange interactions
$J_{i}$, $J_{ci}$ and $J_{c^{\prime }i}\ $represent the $i$%
th-nearest-neighbor interactions between Fe ions within kagome layers, on
the nearest kagome layers, and on the next nearest kagome layers
respectively. As shown in table \ref{JDM}, the in-plane nearest neighbor
coupling $J_{1}$\ favors FM order and is estimated to be -41.97 meV, which
has the similar value with the one in kagome FeSn (around -50 meV) \cite%
{mag-03,mag-11,FeSn-03,FeSn-04}. Note that the distance in $J_{1}$ is 2.5 \r{%
A}\ while the others are all greater than 4 \r{A}. Though there are also AFM
in-plane magnetic interactions such as in-plane next-nearest neighbor
coupling $J_{2}$, they are at least an order of magnitude smaller than $J_{1}
$, resulting in each FM kagome layer. As the out-of-plane nearest neighbor
coupling, $J_{c1}$ is estimated to be 8.44 meV. It makes the magnetic moments stacked
antiferromagnetically between kagome layers, consequently resulting in the
A-type AFM order in kagome\ FeGe, which is consistent with the experiment
\cite{FeGe-1972}. It is worth mentioning that, SOC always exists and leads
to the DM interactions even in the centrosymmetric compound\ FeGe, since not
all Fe-Fe bonds have inversion symmetry. For the equivalent DM interactions
connected by the crystal symmetry (see table \ref{relation1}-\ref{relation3}
in\ Appendix), we only present one of them as a representative. As shown in
table \ref{JDM}, the in-plane nearest neighbor $\mathbf{D}_{1}$ has the form
of (0, 0, $D_{1}^{z}$) according to the crystal symmetry, and $D_{1}^{z}$ is
estimated to be 0.03 meV. Meanwhile, the in-plane next nearest neighbor $%
\mathbf{D}_{2}$ is estimated to be (0, 0, $-$0.12) meV. For the out-of-plane
nearest neighbor, $\mathbf{D}_{c1}$ is zero because its bond has an
inversion center. The other calculated DM interactions are also listed in
table \ref{JDM}, and most of them are small in the order of 0.01 meV.

To explore the magnetic anisotropy in kagome FeGe, we consider the MAE with
the expression $E_{MAE}=K_{2}\sin ^{2}\theta +K_{4}\sin ^{4}\theta $ \cite%
{FeGe-1972,FeGe-1978,FeGe-1984,FeGe-1988}\ neglecting terms of order higher
than four, where $\theta $\ is the angle between the magnetic moment and the
z-axis. The values of $K_{2}$ and $K_{4}$\ are estimated to be 0.066 meV and
0.018 meV respectively based on the approach of Ref. \cite{force-1,force-2},
which are in reasonable agreement with the experimental values 0.021 meV
\cite{FeGe-1988} and 0.012 meV \cite{FeGe-1972}. Here $K_{2}$ and $K_{4}$\
are both positive, making out-of-plane magnetization favored, which is
different from the easy-plane anisotropy in FeSn \cite{mag-11}. Note that
positive $K_{4}$ is the requirement for the stability of the double cone
magnetic structure, which will be discussed below.

According to the experiments \cite{FeGe-1972}, the Curie-Weiss temperature $%
\theta _{CW}$ in kagome\ FeGe is -200 K while the N\'{e}el temperature $%
T_{N} $ is 410 K. The relative low value of the frustration index $|\theta
_{CW}|$/$T_{N}$ (smaller than 1) reveals the interplay of the FM and AFM
interactions \cite{baral2017synthesis}. As shown in table \ref{JDM}, our
calculated results of spin exchange couplings also verify the coexistence of
the FM and AFM interactions. Based on these calculated spin exchange
parameters, we calculate N\'{e}el temperature and Curie-Weiss temperature by
MC simulations \cite{metropolis1949monte,MC-1,MC-2}. The $\theta _{CW}$ and$%
\ T_{N}$ are calculated to be -219 K and 370 K respectively, which agrees
well with the experiment \cite{FeGe-1972}.

\begin{figure}[tbph]
\centering\includegraphics[width=0.48\textwidth]{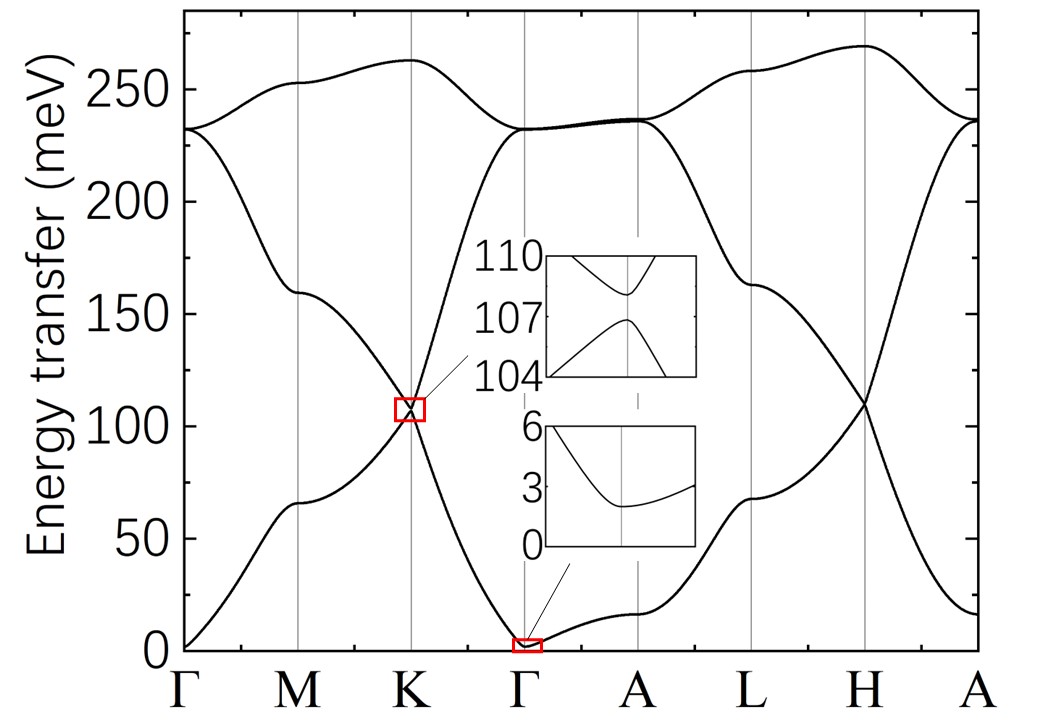}
\caption{Calculated spin-wave dispersion curves along the high-symmetry axis
for FeGe. The insets show the spin gap at $\Gamma$ point induced by
easy-axis anisotropy, and the gap located at about 107 meV of K point
induced by DM interactions.}
\label{spinwave}
\end{figure}

Similar to the electronic structure of a kagome lattice, the spin wave for a
localized spin model with FM nearest-neighbor magnetic exchange also yields
a flat magnetic band and a Dirac magnon \cite{kagomemagnon}. Using the
calculated spin model parameters, one can obtain the magnon spectrum \cite%
{holstein1940field,wang2021determination}.\ The calculated spin-wave
dispersion along the high-symmetry axis is shown in Fig. \ref{spinwave},
which basically captures the key features of kagome lattice geometry.
Similar with FeSn case \cite{mag-03,mag-11,FeSn-03,FeSn-04}, strongly
dispersive magnons in the xy-plane extend to about 260 meV, where the magnon
dispersion along the out-of-plane direction has relatively small bandwidth
of less than 15 meV, reflecting the quasi-two-dimensional magnetic
properties in kagome FeGe. Meanwhile, the Dirac-like node appears at the K
point at about 107 meV, and we find that DM interactions introduce a gap
around 1 meV at the Dirac point, as shown in the inset of Fig. \ref{spinwave}%
. Furthermore, the single-ion anisotropy produces a spin gap of about 2 meV,
which could be verified in future inelastic neutron scattering experiments.

\subsection{The double cone magnetic structure}

\begin{table*}[tbp]
\caption{Four types of $2\times 2\times 2$ CDW phases which can lead to
non-zero DM contribution to double cone spin structure. The corresponding
Wyckoff positions and the coordinates of the atoms in the pristine phase and
these four CDW phases are summarized. }
\label{cdw}%
\begin{tabular}{ccccccccccccccc}
\hline\hline
\multicolumn{3}{c|}{Pristine phase(P6/mmm)} & \multicolumn{3}{c|}{$P622$%
(type \uppercase\expandafter{\romannumeral1})} & \multicolumn{3}{c|}{$P622$%
(type \uppercase\expandafter{\romannumeral2})} & \multicolumn{3}{c|}{$P6$$%
_{3}$$22$(type \uppercase\expandafter{\romannumeral1})} & \multicolumn{3}{c}{%
$P6$$_{3}$$22$(type \uppercase\expandafter{\romannumeral2})} \\ \hline
& WP & \multicolumn{1}{c|}{Coordinates} &  & WP & \multicolumn{1}{c|}{
Coordinates} &  & WP & \multicolumn{1}{c|}{Coordinates} &  & WP &
\multicolumn{1}{c|}{Coordinates} &  & WP & Coordinates \\ \hline
\multirow{4}{*}{Ge1} & \multirow{4}{*}{1a} & \multicolumn{1}{c|}{%
\multirow{4}{*}{(0, 0, 0)}} & Ge1 & 1a & \multicolumn{1}{c|}{(0, 0, 0)} & Ge1
& 2e & \multicolumn{1}{c|}{(0, 0, z$_{1}$)} & Ge1 & 2a & \multicolumn{1}{c|}{
(0, 0, 0)} & Ge1 & 2b & (0, 0, 1/4) \\
&  & \multicolumn{1}{c|}{} & Ge2 & 1b & \multicolumn{1}{c|}{(0, 0, 1/2)} &
Ge2 & 6i & \multicolumn{1}{c|}{(1/2, 0, z$_{2}$)} & Ge2 & 6g &
\multicolumn{1}{c|}{(x$_{1}$, 0, 0)} & Ge2 & 6h & (x$_{1}$, 2x$_{1}$, 1/4)
\\
&  & \multicolumn{1}{c|}{} & Ge3 & 3f & \multicolumn{1}{c|}{(0, 1/2, 0)} &
&  & \multicolumn{1}{c|}{} &  &  & \multicolumn{1}{c|}{} &  &  &  \\
&  & \multicolumn{1}{c|}{} & Ge4 & 3g & \multicolumn{1}{c|}{(0, 1/2, 1/2)} &
&  & \multicolumn{1}{c|}{} &  &  & \multicolumn{1}{c|}{} &  &  &  \\ \hline
\multirow{4}{*}{Ge2} & \multirow{4}{*}{2d} & \multicolumn{1}{c|}{%
\multirow{4}{*}{(1/3, 2/3, 1/2)}} & Ge5 & 4h & \multicolumn{1}{c|}{(1/3,
2/3, z$_{1}$)} & Ge3 & 2c & \multicolumn{1}{c|}{(1/3, 2/3, 0)} & Ge3 & 2c &
\multicolumn{1}{c|}{(1/3, 2/3, 1/4)} & Ge3 & 4f & (1/3, 2/3, z$_{2}$) \\
&  & \multicolumn{1}{c|}{} & Ge6 & 12n & \multicolumn{1}{c|}{(x$_{2}$, y$%
_{2} $, z$_{2}$)} & Ge4 & 2d & \multicolumn{1}{c|}{(1/3, 2/3,1/2)} & Ge4 & 2d
& \multicolumn{1}{c|}{(1/3, 2/3, 3/4)} & Ge4 & 12i & (x$_{3}$, y$_{3}$, z$%
_{3}$) \\
&  & \multicolumn{1}{c|}{} &  &  & \multicolumn{1}{c|}{} & Ge5 & 6l &
\multicolumn{1}{c|}{(x$_{3}$, 2x$_{3}$, 0)} & Ge5 & 6h & \multicolumn{1}{c|}{
(x$_{2}$, 2x$_{2}$, 1/4)} &  &  &  \\
&  & \multicolumn{1}{c|}{} &  &  & \multicolumn{1}{c|}{} & Ge6 & 6m &
\multicolumn{1}{c|}{(x$_{4}$, 2x$_{4}$, 1/2)} & Ge6 & 6h &
\multicolumn{1}{c|}{(x$_{3}$, 2x$_{3}$, 1/4)} &  &  &  \\ \hline
\multirow{4}{*}{Fe} & \multirow{4}{*}{3f} & \multicolumn{1}{c|}{%
\multirow{4}{*}{(1/2, 0, 0)}} & Fe1 & 6j & \multicolumn{1}{c|}{(x$_{3}$, 0,
0)} & Fe1 & 12n & \multicolumn{1}{c|}{(x$_{5}$, y$_{5}$, z$_{5}$)} & Fe1 & 6g
& \multicolumn{1}{c|}{(x$_{4}$, 0, 0)} & Fe1 & 6h & (x$_{4}$, 2x$_{4}$, 1/4)
\\
&  & \multicolumn{1}{c|}{} & Fe2 & 6k & \multicolumn{1}{c|}{(x$_{4}$, 0, 1/2)
} & Fe2 & 12n & \multicolumn{1}{c|}{(x$_{6}$, y$_{6}$, z$_{6}$)} & Fe2 & 6g
& \multicolumn{1}{c|}{(x$_{5}$, 0, 0)} & Fe2 & 6h & (x$_{5}$, 2x$_{5}$, 1/4)
\\
&  & \multicolumn{1}{c|}{} & Fe3 & 6l & \multicolumn{1}{c|}{(x$_{5}$, 2x$%
_{5} $, 0)} &  &  & \multicolumn{1}{c|}{} & Fe3 & 12i & \multicolumn{1}{c|}{
(x$_{6}$, y$_{6}$, z$_{6}$)} & Fe3 & 12i & (x$_{6}$, y$_{6}$, z$_{6}$) \\
&  & \multicolumn{1}{c|}{} & Fe4 & 6m & \multicolumn{1}{c|}{(x$_{6}$, 2x$%
_{6} $, 1/2)} &  &  & \multicolumn{1}{c|}{} &  &  & \multicolumn{1}{c|}{} &
&  &  \\ \hline\hline
\end{tabular}%
\end{table*}

At $T_{canting}$ = 60 K, the kagome lattice FeGe becomes
a c-axis double cone AFM structure \cite%
{FeGe-1972,FeGe-1975,FeGe-1978,FeGe-1984,FeGe-1988} where the magnetic
ground state could be written as Eq. (\ref{cone}) in Appendix. Considering
the magnetic interactions and the MAE, the total energy of the double cone
spin structure could be written as Eq. (\ref{etot}) in Appendix. When DM
interactions are not considered, the extremum condition of the total energy
gives the equilibrium value of wave vector $\delta $ and the cone half angle
$\theta $ (i.e. Eq. (\ref{cosq}) and (\ref{sin}) in Appendix)

\begin{eqnarray}
\cos \delta &=&\frac{\sum_{i}N_{ci}J_{ci}}{4\sum_{i}N_{c^{\prime
}i}J_{c^{\prime }i}} \\
\sin ^{2}\theta &=&-\frac{K_{2}-\frac{1}{2N}\sum_{i}N_{c^{\prime
}i}J_{c^{\prime }i}\delta ^{4}}{2K_{4}}  \label{mainsin}
\end{eqnarray}%
\ \

Note that the minimum of the total energy requires that the second
derivative of Eq. (\ref{etot}) in Appendix is positive, thus $K_{4}$ must be positive.
Hence $K_{2}-\frac{1}{2N}\sum_{i}N_{c^{\prime }i}J_{c^{\prime }i}\delta ^{4}$
(i.e. the numerator\ of Eq. (\ref{mainsin})) must be negative. However, our
reasonable magnetic parameters cannot explain the double cone magnetic
ground state. The value of wave vector $\delta $ is small in experimental
measurement (0.17 in Ref. \cite{FeGe-1972} and 0.25 in Ref. \cite{FeGe-1977}%
), thus $\delta ^{4}$ is around 0.001. Meanwhile, the value of $\frac{1}{2N}%
\sum_{i}N_{c^{\prime }i}J_{c^{\prime }i}$ is of the order of 1 meV, which obviously cannot explain the double cone magnetic
structure \cite{FeGe-1972}.

We thus consider the effect of DM interactions on double cone spin
structure. Since the exchange interactions between two next nearest neighbor
kagome layers are relatively small, we only consider the Heisenberg and DM
interactions between two nearest neighbor kagome layers, i.e. $J_{ci}$ and $%
\mathbf{D}_{ci}$. We find that wave vector $\delta $ and the cone half angle
$\theta $ have the expressions as (i.e. Eq. (\ref{tanq}) and (\ref{sin2}) in
Appendix)%
\begin{eqnarray}
\tan \delta  &=&\frac{\sum_{i,j}D_{ci,j}^{z}}{\sum_{i}N_{ci}J_{ci}} \\
\sin ^{2}\theta  &=&-\frac{K_{2}-\frac{1}{2N}\sum_{i,j}D_{ci,j}^{z}\delta }{%
2K_{4}}  \label{mainsin2}
\end{eqnarray}

It should be noted that, comparing Eq. (\ref{mainsin}) and (\ref{mainsin2}),
DM interactions are more efficient than Heisenberg interactions for causing
double cone spin structure since $\delta $ is small. Though the space group $%
P6/mmm$ of high-temperature phase in FeGe has a global inversion center, not
all Fe-Fe bonds have inversion symmetry and DM interactions could exist.
However, according to the inversion symmetry of space group $P6/mmm$, the
total contribution of DM interactions to the energy of double cone magnetic
structure in Eq. (\ref{etot}) is absent, i.e. $\sum_{i,j}D_{ci,j}^{z}=0$
(see Appendix D). Meanwhile, mirror symmetries in space group $P6/mmm$ would
also eliminate the contribution of DM interactions based on the symmetry
analysis. Therefore, DM interactions have no net contribution to double cone
magnetic structure with the symmetry of high-temperature phase. For the CDW
phases with the space group of $P6/mmm$ suggested by Ref. \cite%
{2206.12033,2210.06359} (the first two structures of the table \ref{cdw1} in
Appendix), the total contribution of DM interactions is still absent and
cannot explain the magnetic ground state of double cone spin structure.

\subsection{The interplay of CDW and double cone structure}

As mentioned above, DM interactions play a more important role in the
double cone spin structure. Meanwhile, it is very sensitive to atomic
displacements. Therefore, in the following we explore the CDW phases with
symmetry-allowed DM contribution to double cone spin structure which may
explain the canted magnetic ground state.

The $2\times 2\times 2$ supercell structure of CDW phase (compared with the
nonmagnetic pristine phase) is suggested experimentally \cite%
{teng2022discovery,yin2022discovery,2206.12033,2210.06359}. Considering all
CDW\ phases whose associated point group in the maximal subgroups of $D_{6h}$%
, we find 68 different possible CDW phases which are the subgroups of the
parent $P6/mmm$ phase with $2\times 2\times 2$ supercell (see details in
Appendix D). The corresponding relations of atomic positions between the
pristine phase and these proposed CDW phases are all summarized in table \ref%
{cdw1}-\ref{cdw5} of Appendix.

Note that the inversion symmetry and mirror symmetries would all eliminate the net contribution of DM interactions as disscussed
above. We find that among these 68 proposed CDW phases, only four distorted
structures break all these symetries above, and can lead to non-zero DM
contribution in Eq. (\ref{etot}) of Appendix. We list the corresponding
Wyckoff positions (WP) and the coordinates of the atoms in the pristine
phase and these four CDW phases in table \ref{cdw}. They comes from two
space groups $P622$ and $P6_{3}22$. It should be mentioned that there are
two different CDW phases for each of these two space groups, which are
labeled as (type I) and (type II) in table \ref{cdw}. Note that the CDW
phase with $P622$ space group is also suggested in Ref. \cite%
{teng2022discovery}.

Raman spectroscopy is a fast and usually non-destructive technique which can
be used to characterize the structural distortion of materials. Based on the
atomic coordinates in table \ref{cdw}, we predict the irreducible
representation of the Raman active modes of these four proposed CDW phases
using symmetry analysis \cite{Bradley}. For $P622$ (type I) and (type II) CDW phases, the
Raman active modes are 8$\mathrm{{A_{1}}}$ + 26$\mathrm{{E_{1}}}$ + 22$%
\mathrm{{E_{2}}}$ and 10$\mathrm{{A_{1}}}$ + 26$\mathrm{{E_{1}}}$ + 22$%
\mathrm{{E_{2}}}$. Meanwhile, for $P6_{3}22$ (type I) and
(type II), the Raman active modes are 8$\mathrm{{A_{1}}}$ + 24$\mathrm{{E_{1}%
}}$ + 24$\mathrm{{E_{2}}}$ and 10$\mathrm{{A_{1}}}$ + 24$\mathrm{{E_{1}}}$ +
24$\mathrm{{E_{2}}}$, respectively. Note that even within the
same symmetry of space group $P622$, the different structural distortion of
CDW phases $P622$ (type I) and (type II) could result in the different
number of Raman active modes (56 and 58 respectively), which could be
identified by Raman spectroscopy.

\section{CONCLUSION}

In conclusion, we systematically analyze the electronic and magnetic
properties of kagome FeGe. Our numerical results show that this material is
a magnetic metal exhibiting large magnetic splitting around 1.8 eV. The
magnetic splitting makes the flat bands away from Fermi level, and bring two
vHSs near the Fermi level. We estimate the magnetic parameters, and find
that the ferromagnetic nearest-neighbor $J_{1}$ dominates over the others,
while the magnetic interactions between nearest kagome layers favors
antiferromagnetic. Based on these spin exchange parameters, the calculated N%
\'{e}el temperature and Curie-Weiss temperature also agree well with the
experiments. Furthermore, the magnetic excitation spectra are calculated
using linear spin wave theory and a spin gap about 2 meV is predicted. Note
that the double cone magnetic transition at a lower temperature cannot be
reproduced by these reasonable magnetic parameters. Meanwhile, due to the
inversion symmetry and mirror symmetries in the space group $P6/mmm$
of high-temperature phase, the total contribution of DM interactions to the
double cone magnetic structure is absent. Since DM interactions are very
sensitive to small atomic displacements and symmetry restrictions, and also much more
efficient than Heisenberg interactions for causing this canted spin
structure, we propose that the double cone spin structure may arise from the
structural distortion. We explore 68 possible CDW phases of kagome FeGe
which are subgroups of the pristine phase with $2\times 2\times 2$
supercell, and symmetry-allowed four CDW structures which have non-zero DM
contribution and may result in double cone spin structure are proposed.
These four CDW phases belong to two space groups $P622$ and $P6_{3}22$, and
we further propose that they can be identified from their different numbers
of Raman active peaks. Therefore, we believe that symmetry analysis
plays an important role in exploring the possible structural
distortion in complex magnetic configurations.

\section{Acknowledgements}

This work was supported by the NSFC (No. 12188101, 11834006, 12004170,
11790311, 51721001), National Key R\&D Program of China (No.
2018YFA0305704), Natural Science Foundation of Jiangsu Province, China
(Grant No. BK20200326), and the excellent programme in Nanjing University.
Xiangang Wan also acknowledges the support from the Tencent Foundation
through the XPLORER PRIZE.

\section{Appendix}

\subsection{The density of states in kagome FeGe}

\begin{figure}[tbph]
\centering\includegraphics[width=0.48\textwidth]{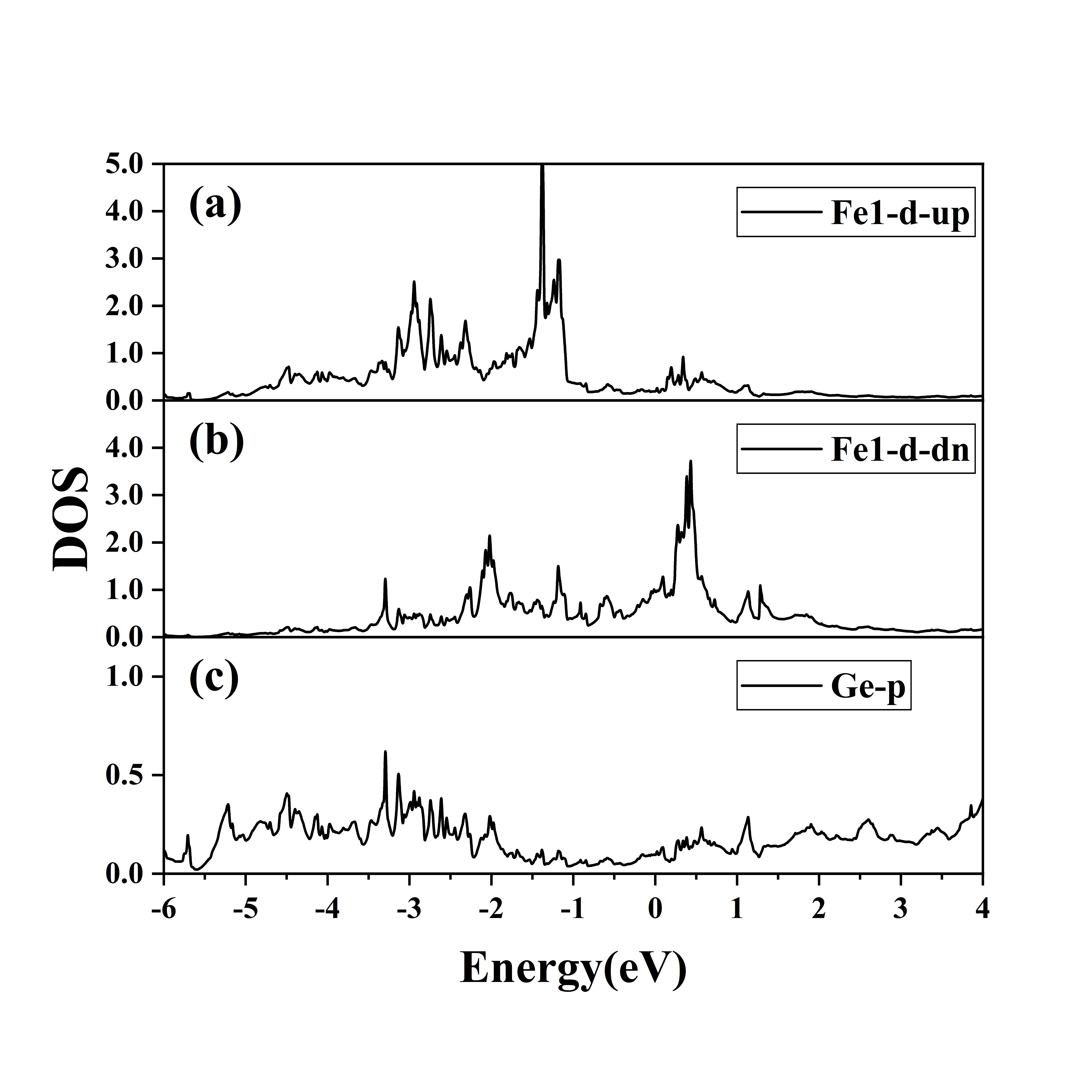}
\caption{Partial DOS of FeGe from LSDA + SOC
calculations. The Fermi energy is set to zero. (a) and (b) represent the
spin-up and spin-down channel of $d$ orbitals in Fe1 atom located at
(1/2,0,0), while (c) represents the DOS of Ge-$p$ orbitals.}
\label{dos}
\end{figure}

The Partial density of states (DOS) of FeGe from LSDA + SOC
calculations are shown in Fig. \ref{dos}.

\subsection{The symmetry restrictions on the magnetic interactions}

\begin{table}[tbh]
\caption{The distances, the bond information and the symmetry restricted
interactions of corresponding Fe ions within xy-planes. Here $n$, $n^{\prime
}$ and $R_{l}$\ correspond to $\mathbf{J}_{\boldsymbol{\protect\tau }_{n},%
\boldsymbol{\protect\tau _{n^{\prime }}+}\mathbf{R}_{l}}$, where $R_{l}$ and
$\protect\tau _{n}$ represent the lattice translation vector and the
position of magnetic ions in the lattice basis.\ Three magnetic ions are
located at $\protect\tau _{1}$ (1/2, 0, 0), $\protect\tau _{2}$ (0, 1/2, 0),
and $\protect\tau _{3}$ (1/2, 1/2, 0). The equivalent $\mathbf{D}_{i}$'s are
labeled as the sub-index of $j$, i.e. the $\mathbf{D}_{i,j}$ in the table.}
\label{relation1}%
\begin{tabular}{c|ccc|cc}
\hline\hline
Distance($\mathring{\mathrm{A}}$) & $n$ & $n^{\prime }$ & $R_{l}$ & $J$ & DM
\\ \hline
2.50 & 3 & 1 & (0,1,0) & $J_{1}$ & $\mathbf{D}_{1,1}(0,0,D_{1}^{z})$ \\
& 1 & 2 & (0,-1,0) & $J_{1}$ & $\mathbf{D}_{1,2}(0,0,D_{1}^{z})$ \\
& 2 & 3 & (0,0,0) & $J_{1}$ & $\mathbf{D}_{1,3}(0,0,D_{1}^{z})$ \\
& 3 & 1 & (0,0,0) & $J_{1}$ & $\mathbf{D}_{1,4}(0,0,D_{1}^{z})$ \\
& 1 & 2 & (1,0,0) & $J_{1}$ & $\mathbf{D}_{1,5}(0,0,D_{1}^{z})$ \\
& 2 & 3 & (-1,0,0) & $J_{1}$ & $\mathbf{D}_{1,6}(0,0,D_{1}^{z})$ \\ \hline
4.33 & 1 & 2 & (1,-1,0) & $J_{2}$ & $\mathbf{D}_{2,1}(0,0,D_{2}^{z})$ \\
& 2 & 3 & (0,1,0) & $J_{2}$ & $\mathbf{D}_{2,2}(0,0,D_{2}^{z})$ \\
& 3 & 1 & (-1,0,0) & $J_{2}$ & $\mathbf{D}_{2,3}(0,0,D_{2}^{z})$ \\
& 1 & 2 & (0,0,0) & $J_{2}$ & $\mathbf{D}_{2,4}(0,0,D_{2}^{z})$ \\
& 2 & 3 & (-1,-1,0) & $J_{2}$ & $\mathbf{D}_{2,5}(0,0,D_{2}^{z})$ \\
& 3 & 1 & (1,1,0) & $J_{2}$ & $\mathbf{D}_{2,6}(0,0,D_{2}^{z})$ \\
\hline\hline
\end{tabular}%
\end{table}

\begin{table}[tbh!]
\caption{The distances, the bond information and the symmetry restricted
interactions of corresponding Fe ions between nearest-neighbor $(001)$%
-planes. Here $n$, $n^{\prime }$ and $R_{l}$\ correspond to $\mathbf{J}_{%
\boldsymbol{\protect\tau }_{n},\boldsymbol{\protect\tau _{n^{\prime }}+}%
\mathbf{R}_{l}}$, where $R_{l}$ and $\protect\tau _{n}$ represent the
lattice translation vector and the position of magnetic ions in the lattice
basis. Three magnetic ions are located at $\protect\tau _{1}$ (1/2, 0, 0), $%
\protect\tau _{2}$ (0, 1/2, 0), and $\protect\tau _{3}$ (1/2, 1/2, 0). The
equivalent $\mathbf{D}_{ci}$'s are labeled as the sub-index of $j$, i.e. the
$\mathbf{D}_{ci,j}$ in the table.}
\label{relation2}%
\begin{tabular}{c|ccc|cc}
\hline\hline
Distance($\mathring{\mathrm{A}}$) & $n$ & $n^{\prime }$ & $R_{l}$ & $J$ & DM
\\ \hline
4.05 & 1 & 1 & (0,0,1) & $J_{c1}$ & $\mathbf{D}_{c1,1}(0,0,0)$ \\
& 2 & 2 & (0,0,1) & $J_{c1}$ & $\mathbf{D}_{c1,2}(0,0,0)$ \\
& 3 & 3 & (0,0,1) & $J_{c1}$ & $\mathbf{D}_{c1,3}(0,0,0)$ \\ \hline
4.76 & 3 & 1 & (0,0,1) & $J_{c2}$ & $\mathbf{D}_{c2,1}(D_{c2}^{x},-\sqrt{3}%
D_{c2}^{x},D_{c2}^{z})$ \\
& 1 & 2 & (1,0,1) & $J_{c2}$ & $\mathbf{D}_{c2,2}(D_{c2}^{x},\sqrt{3}%
D_{c2}^{x},D_{c2}^{z})$ \\
& 2 & 3 & (-1,0,1) & $J_{c2}$ & $\mathbf{D}%
_{c2,3}(-2D_{c2}^{x},0,D_{c2}^{z}) $ \\
& 3 & 1 & (0,1,1) & $J_{c2}$ & $\mathbf{D}_{c2,4}(-D_{c2}^{x},\sqrt{3}%
D_{c2}^{x},D_{c2}^{z})$ \\
& 1 & 2 & (0,-1,1) & $J_{c2}$ & $\mathbf{D}_{c2,5}(-D_{c2}^{x},-\sqrt{3}%
D_{c2}^{x},D_{c2}^{z}) $ \\
& 2 & 3 & (0,0,1) & $J_{c2}$ & $\mathbf{D}_{c2,6}(2D_{c2}^{x},0,D_{c2}^{z})$
\\
& 1 & 3 & (0,-1,1) & $J_{c2}$ & $\mathbf{D}_{c2,7}(-D_{c2}^{x},\sqrt{3}%
D_{c2}^{x},-D_{c2}^{z}) $ \\
& 2 & 1 & (0,1,1) & $J_{c2}$ & $\mathbf{D}_{c2,8}(-D_{c2}^{x},-\sqrt{3}%
D_{c2}^{x},-D_{c2}^{z}) $ \\
& 3 & 2 & (0,0,1) & $J_{c2}$ & $\mathbf{D}_{c2,9}(2D_{c2}^{x},0,-D_{c2}^{z})$
\\
& 1 & 3 & (0,0,1) & $J_{c2}$ & $\mathbf{D}_{c2,10}(D_{c2}^{x},-\sqrt{3}%
D_{c2}^{x},-D_{c2}^{z})$ \\
& 2 & 1 & (-1,0,1) & $J_{c2}$ & $\mathbf{D}_{c2,11}(D_{c2}^{x},\sqrt{3}%
D_{c2}^{x},-D_{c2}^{z}) $ \\
& 3 & 2 & (1,0,1) & $J_{c2}$ & $\mathbf{D}%
_{c2,12}(-2D_{c2}^{x},0,-D_{c2}^{z})$ \\ \hline
5.93 & 2 & 1 & (-1,1,1) & $J_{c3}$ & $\mathbf{D}_{c3,1}(-\sqrt{3}%
D_{c3}^{y},D_{c3}^{y},-D_{c3}^{z})$ \\
& 3 & 2 & (0,-1,1) & $J_{c3}$ & $\mathbf{D}%
_{c3,2}(0,-2D_{c3}^{y},-D_{c3}^{z})$ \\
& 1 & 3 & (1,0,1) & $J_{c3}$ & $\mathbf{D}_{c3,3}(\sqrt{3}%
D_{c3}^{y},D_{c3}^{y},-D_{c3}^{z})$ \\
& 2 & 1 & (0,0,1) & $J_{c3}$ & $\mathbf{D}_{c3,4}(\sqrt{3}%
D_{c3}^{y},-D_{c3}^{y},-D_{c3}^{z})$ \\
& 3 & 2 & (1,1,1) & $J_{c3}$ & $\mathbf{D}_{c3,5}(0,2D_{c3}^{y},-D_{c3}^{z})$
\\
& 1 & 3 & (-1,-1,1) & $J_{c3}$ & $\mathbf{D}_{c3,6}(-\sqrt{3}%
D_{c3}^{y},-D_{c3}^{y},-D_{c3}^{z})$ \\
& 1 & 2 & (0,0,1) & $J_{c3}$ & $\mathbf{D}_{c3,7}(\sqrt{3}%
D_{c3}^{y},-D_{c3}^{y},D_{c3}^{z})$ \\
& 2 & 3 & (-1,-1,1) & $J_{c3}$ & $\mathbf{D}%
_{c3,8}(0,2D_{c3}^{y},D_{c3}^{z}) $ \\
& 3 & 1 & (1,1,1) & $J_{c3}$ & $\mathbf{D}_{c3,9}(-\sqrt{3}%
D_{c3}^{y},-D_{c3}^{y},D_{c3}^{z})$ \\
& 1 & 2 & (1,-1,1) & $J_{c3}$ & $\mathbf{D}_{c3,10}(-\sqrt{3}%
D_{c3}^{y},D_{c3}^{y},D_{c3}^{z})$ \\
& 2 & 3 & (0,1,1) & $J_{c3}$ & $\mathbf{D}%
_{c3,11}(0,-2D_{c3}^{y},D_{c3}^{z}) $ \\
& 3 & 1 & (-1,0,1) & $J_{c3}$ & $\mathbf{D}_{c3,12}(\sqrt{3}%
D_{c3}^{y},D_{c3}^{y},D_{c3}^{z})$ \\ \hline\hline
\end{tabular}%
\end{table}

\begin{table}[tbh]
\caption{The distances, the bond information and the symmetry restricted
interactions of corresponding Fe ions between next-nearest-neighbor $(001)$%
-planes. Here $n$, $n^{\prime }$ and $R_{l}$\ correspond to $\mathbf{J}_{%
\boldsymbol{\protect\tau }_{n},\boldsymbol{\protect\tau _{n^{\prime }}+}%
\mathbf{R}_{l}}$, where $R_{l}$ and $\protect\tau _{n}$ represent the
lattice translation vector and the position of magnetic ions in the lattice
basis.\ Three magnetic ions are located at $\protect\tau _{1}$ (1/2, 0, 0), $%
\protect\tau _{2}$ (0, 1/2, 0), and $\protect\tau _{3}$ (1/2, 1/2, 0). The
equivalent $\mathbf{D}_{c^{\prime }i}$'s are labeled as the sub-index of $j$%
, i.e. the $\mathbf{D}_{c^{\prime }i,j}$ in the table.}
\label{relation3}%
\begin{tabular}{c|ccc|cc}
\hline\hline
Distance($\mathring{\mathrm{A}}$) & $n$ & $n^{\prime }$ & $R_{l}$ & $J$ & DM
\\ \hline
8.11 & 1 & 1 & (0,0,2) & $J_{c^{\prime }1}$ & $\mathbf{D}_{c^{\prime
}1,1}(0,0,0)$ \\
& 2 & 2 & (0,0,2) & $J_{c^{\prime }1}$ & $\mathbf{D}_{c^{\prime }1,2}(0,0,0)$
\\
& 3 & 3 & (0,0,2) & $J_{c^{\prime }1}$ & $\mathbf{D}_{c^{\prime }1,3}(0,0,0)$
\\ \hline
8.49 & 2 & 1 & (0,1,2) & $J_{c^{\prime }2}$ & $\mathbf{D}_{c^{\prime
}2,1}(D_{c^{\prime }2}^{x},\sqrt{3}D_{c^{\prime }2}^{x},D_{c^{\prime
}2}^{z}) $ \\
& 3 & 2 & (0,0,2) & $J_{c^{\prime }2}$ & $\mathbf{D}_{c^{\prime
}2,2}(-2D_{c^{\prime }2}^{x},0,D_{c^{\prime }2}^{z})$ \\
& 1 & 3 & (0,-1,2) & $J_{c^{\prime }2}$ & $\mathbf{D}_{c^{\prime
}2,3}(D_{c^{\prime }2}^{x},-\sqrt{3}D_{c^{\prime }2}^{x},D_{c^{\prime
}2}^{z})$ \\
& 2 & 1 & (-1,0,2) & $J_{c^{\prime }2}$ & $\mathbf{D}_{c^{\prime
}2,4}(-D_{c^{\prime }2}^{x},-\sqrt{3}D_{c^{\prime }2}^{x},D_{c^{\prime
}2}^{z})$ \\
& 3 & 2 & (1,0,2) & $J_{c^{\prime }2}$ & $\mathbf{D}_{c^{\prime
}2,5}(2D_{c^{\prime }2}^{x},0,D_{c^{\prime }2}^{z})$ \\
& 1 & 3 & (0,0,2) & $J_{c^{\prime }2}$ & $\mathbf{D}_{c^{\prime
}2,6}(-D_{c^{\prime }2}^{x},\sqrt{3}D_{c^{\prime }2}^{x},D_{c^{\prime
}2}^{z})$ \\
& 1 & 2 & (1,0,2) & $J_{c^{\prime }2}$ & $\mathbf{D}_{c^{\prime
}2,7}(-D_{c^{\prime }2}^{x},-\sqrt{3}D_{c^{\prime }2}^{x},-D_{c^{\prime
}2}^{z})$ \\
& 2 & 3 & (-1,0,2) & $J_{c^{\prime }2}$ & $\mathbf{D}_{c^{\prime
}2,8}(2D_{c^{\prime }2}^{x},0,-D_{c^{\prime }2}^{z})$ \\
& 3 & 1 & (0,0,2) & $J_{c^{\prime }2}$ & $\mathbf{D}_{c^{\prime
}2,9}(-D_{c^{\prime }2}^{x},\sqrt{3}D_{c^{\prime }2}^{x},-D_{c^{\prime
}2}^{z})$ \\
& 1 & 2 & (0,-1,2) & $J_{c^{\prime }2}$ & $\mathbf{D}_{c^{\prime
}2,10}(D_{c^{\prime }2}^{x},\sqrt{3}D_{c^{\prime }2}^{x},-D_{c^{\prime
}2}^{z})$ \\
& 2 & 3 & (0,0,2) & $J_{c^{\prime }2}$ & $\mathbf{D}_{c^{\prime
}2,11}(-2D_{c^{\prime }2}^{x},0,-D_{c^{\prime }2}^{z})$ \\
& 3 & 1 & (0,1,2) & $J_{c^{\prime }2}$ & $\mathbf{D}_{c^{\prime
}2,12}(D_{c^{\prime }2}^{x},-\sqrt{3}D_{c^{\prime }2}^{x},-D_{c^{\prime
}2}^{z})$ \\ \hline\hline
\end{tabular}%
\end{table}

Here we consider a general pairwise spin model

\begin{equation}
H=\sum_{l,n,l^{\prime },n^{\prime }}\mathbf{S}_{ln}\mathbf{J}_{\mathbf{R}%
_{l}+\boldsymbol{\tau }_{n},\mathbf{R}_{l^{\prime }}+\boldsymbol{\tau
_{n^{\prime }}}}\mathbf{S}_{l^{\prime }n^{\prime }}  \label{spinmodel}
\end{equation}

where $\mathbf{J}_{\mathbf{R}_{l}+\boldsymbol{\tau }_{n},\mathbf{R}%
_{l^{\prime }}+\boldsymbol{\tau _{n^{\prime }}}}$, a $3\times 3$ tensor,
represents the spin exchange parameters. $\mathbf{R}_{l}$ and $\boldsymbol{%
\tau} _{n}$ represent the lattice translation vector and the position of
magnetic ions in the lattice basis, and $\mathbf{S}_{ln}$ means the spin at
the site of $\mathbf{R}_{l}+\boldsymbol{\tau }_{n}.$Translation symmetry
will restrict $\mathbf{J}_{\mathbf{R}_{l}+\boldsymbol{\tau }_{n},\mathbf{R}%
_{l^{\prime }}+\boldsymbol{\tau _{n^{\prime }}}}$ to be only related to $%
\mathbf{J}_{\boldsymbol{\tau }_{n},\boldsymbol{\tau _{n^{\prime }}+}\mathbf{R%
}_{l^{\prime \prime }}}$ where $R_{l^{\prime \prime }}=R_{l^{\prime }}-R_{l}$%
, irrespective of the starting unit cell. Other spatial symmetries will also
give restrictions on the magnetic exchange interactions. We consider a
general space group element $\{\alpha |\mathbf{t}\}$, where the left part
represents the rotation and the right part means the lattice translation.
Supposing under this symmetry operator, $\mathbf{R}_{m}+\boldsymbol{\tau }%
_{p}$ and $\mathbf{R}_{m^{\prime }}+\boldsymbol{\tau _{p^{\prime }}}$
transfer to $\mathbf{R}_{l}+\boldsymbol{\tau }_{n}$ and $\mathbf{R}%
_{l^{\prime }}+\boldsymbol{\tau _{n^{\prime }}}$, respectively, meanwhile
the transformation of spin becomes $\mathbf{S}_{mp}=M(\alpha )\mathbf{S}%
_{ln} $, where $M(\alpha )$ is the representation matrix of the proper
rotation part of the operation $\alpha $ in the coordinate system, we get
the following expression:

\begin{eqnarray}
H &=&\sum_{l,n,l^{\prime },n^{\prime }}\mathbf{S}_{ln}\mathbf{J}_{\mathbf{R}%
_{l}+\boldsymbol{\tau }_{n},\mathbf{R}_{l^{\prime }}+\boldsymbol{\tau
_{n^{\prime }}}}\mathbf{S}_{l^{\prime }n^{\prime }}  \notag \\
&=&\sum_{l,n,l^{\prime },n^{\prime }}\mathbf{S}_{ln}M^{\dag }(\alpha
)M(\alpha )\mathbf{J}_{\mathbf{R}_{l}+\boldsymbol{\tau }_{n},\mathbf{R}%
_{l^{\prime }}+\boldsymbol{\tau _{n^{\prime }}}}M^{\dag }(\alpha )M(\alpha )%
\mathbf{S}_{l^{\prime }n^{\prime }}  \notag \\
&=&\sum_{m,p,m^{\prime },p^{\prime }}\mathbf{S}_{mp}\left[ M(\alpha )\mathbf{%
J}_{\mathbf{R}_{l}+\boldsymbol{\tau }_{n},\mathbf{R}_{l^{\prime }}+%
\boldsymbol{\tau _{n^{\prime }}}}M^{\dag }(\alpha )\right] \mathbf{S}%
_{m^{\prime }p^{\prime }}
\end{eqnarray}

Then the exchange interactions should satisfy the following condition:

\begin{equation}
\mathbf{J}_{\mathbf{R}_{m}+\boldsymbol{\tau }_{p},\mathbf{R}_{m^{\prime }}+%
\boldsymbol{\tau _{p^{\prime }}}}=M(\alpha )\mathbf{J}_{\mathbf{R}_{l}+%
\boldsymbol{\tau }_{n},\mathbf{R}_{l^{\prime }}+\boldsymbol{\tau _{n^{\prime
}}}}M^{\dag }(\alpha )  \label{Jrelation}
\end{equation}

After decomposing the $3\times 3$ tensor $\mathbf{J}$ into scalar Heisenberg
term $J$ and vector DM term $\mathbf{D}$ as in the maintext, we obtain the
following results:

\begin{eqnarray}
J_{\mathbf{R}_{m}+\boldsymbol{\tau }_{p},\mathbf{R}_{m^{\prime }}+%
\boldsymbol{\tau _{p^{\prime }}}} &=&J_{\mathbf{R}_{l}+\boldsymbol{\tau }%
_{n},\mathbf{R}_{l^{\prime }}+\boldsymbol{\tau _{n^{\prime }}}}  \notag \\
\mathbf{D}_{\mathbf{R}_{m}+\boldsymbol{\tau }_{p},\mathbf{R}_{m^{\prime }}+%
\boldsymbol{\tau _{p^{\prime }}}} &=&M(\alpha )\mathbf{D}_{\mathbf{R}_{l}+%
\boldsymbol{\tau }_{n},\mathbf{R}_{l^{\prime }}+\boldsymbol{\tau _{n^{\prime
}}}}  \label{Jrelation-2}
\end{eqnarray}

Meanwhile, it is should be noted that the Heisenberg and DM interactions
obey the following commutation relations

\begin{eqnarray}
J_{\mathbf{R}_{l^{\prime }}+\boldsymbol{\tau _{n^{\prime }},}\mathbf{R}_{l}+%
\boldsymbol{\tau }_{n}} &=&J_{\mathbf{R}_{l}+\boldsymbol{\tau }_{n},\mathbf{R%
}_{l^{\prime }}+\boldsymbol{\tau _{n^{\prime }}}}  \notag \\
\mathbf{D}_{\mathbf{R}_{l^{\prime }}+\boldsymbol{\tau _{n^{\prime }},}%
\mathbf{R}_{l}+\boldsymbol{\tau }_{n}} &=&-\mathbf{D}_{\mathbf{R}_{l}+%
\boldsymbol{\tau }_{n},\mathbf{R}_{l^{\prime }}+\boldsymbol{\tau _{n^{\prime
}}}}  \label{Jrelation-3}
\end{eqnarray}

According to the above equations (i.e. Eq. (\ref{Jrelation-2}) and (\ref%
{Jrelation-3})), one can obtain the symmetry restricted magnetic
interactions for kagome FeGe with space group $P6/mmm$, as shown in table %
\ref{relation1}, \ref{relation2} and \ref{relation3}. Note that the
equivalent $\mathbf{D}_{i}$'s are labeled as the sub-index of $j$, i.e. the $%
\mathbf{D}_{i,j}$ in table \ref{relation1}, \ref{relation2} and \ref%
{relation3}.

\subsection{The details of double cone structure}

According to the experimental works \cite%
{FeGe-1972,FeGe-1978,FeGe-1984,FeGe-1988}, in hexagonal FeGe there is a
transition from a uniaxial spin system to a double cone spin structure at $%
T_{canting}$ = 60 K \cite{FeGe-1972}, which is expressed by the following
equations:

\begin{eqnarray}
\left\langle S^{x}\right\rangle &=&S\sin \theta \cos \left( \left( \pi \pm
\delta \right) \frac{z}{c}+\varphi \right)  \notag \\
\left\langle S^{y}\right\rangle &=&S\sin \theta \sin \left( \left( \pi \pm
\delta \right) \frac{z}{c}+\varphi \right)  \notag \\
\left\langle S^{z}\right\rangle &=&S\cos \theta \cos \left( \frac{\pi z}{c}%
\right)  \label{cone}
\end{eqnarray}

where $\theta $ is the cone half angle, and $c$ represents the lattice
parameter. If $\delta =0$ there will be a simple tilting of the spins. When $%
\delta $ represents the small angle, Eq. (\ref{cone}) gives a double cone
spin structure. Following the previous works \cite%
{FeGe-1972,FeGe-1978,FeGe-1984,FeGe-1988}, here we consider the MAE with the
expression neglecting terms of order higher than four written as

\begin{equation}
E_{MAE}=K_{2}\sin ^{2}\theta +K_{4}\sin ^{4}\theta   \label{mae}
\end{equation}

Therefore the total energy of Eq. (\ref{spinmodel}) and Eq. (\ref{mae}) in
double cone spin structure per unit cell could be written as

\begin{eqnarray}
E(\delta ,\theta ) &=&\sum_{i}N_{ci}J_{ci}(-\sin ^{2}\theta \cos \delta
-\cos ^{2}\theta )  \notag \\
&&+\sum_{i}N_{c^{\prime }i}J_{c^{\prime }i}(\sin ^{2}\theta \cos 2\delta
+\cos ^{2}\theta )  \notag \\
&&-\sum_{i,j}D_{ci,j}^{z}(\sin ^{2}\theta \sin \delta )  \notag \\
&&-\sum_{i,j}D_{c^{\prime }i,j}^{z}(\sin ^{2}\theta \sin 2\delta )  \notag \\
&&+N(K_{2}\sin ^{2}\theta +K_{4}\sin ^{4}\theta )  \label{etot}
\end{eqnarray}

where $N_{ci}$ and $N_{c^{\prime }i}$\ are the corresponding number of
neighbors of $J_{ci}$ and $J_{c^{\prime }i}$, and $N$ represents the number
of magnetic ions in one unit cell. When DM interactions are not considered,
the extremum condition in total energ gives the equilibrium value of wave
vector $\delta $ with following equation \cite{FeGe-1972,FeGe-1988}:

\begin{equation}
\cos \delta =\frac{\sum_{i}N_{ci}J_{ci}}{4\sum_{i}N_{c^{\prime
}i}J_{c^{\prime }i}}  \label{cosq}
\end{equation}

While the cone half angle $\theta $ has the expression as%
\begin{equation}
\sin ^{2}\theta =-\frac{K_{2}-\frac{1}{2N}\sum_{i}N_{c^{\prime
}i}J_{c^{\prime }i}\delta ^{4}}{2K_{4}}  \label{sin}
\end{equation}%
\ \

A minimum in the total energy (see Eq. (\ref{etot})) will occur only if $%
K_{4}$ is positive, and Eq. (\ref{sin}) requires that $K_{2}-\frac{1}{2N}%
\sum_{i}N_{c^{\prime }i}J_{c^{\prime }i}\delta ^{4}$ must be negative.

When the magnetic interactions including Heisenberg and DM interactions
between two nearest neighbor xy-planes, i.e. $J_{ci}$ and $\mathbf{D}_{ci}$,
are considered, the equilibrium value of wave vector $\delta $ is obtained
by the minimum in total energy written as%
\begin{equation}
\tan \delta =\frac{\sum_{i,j}D_{ci,j}^{z}}{\sum_{i}N_{ci}J_{ci}}
\label{tanq}
\end{equation}

where $j$ is the sub-index of the equivalent $\mathbf{D}_{ci}$'s. Meanwhile,
we find the following expression for $\theta $%
\begin{equation}
\sin ^{2}\theta =-\frac{K_{2}-\frac{1}{2N}\sum_{i,j}D_{ci,j}^{z}\delta }{%
2K_{4}}  \label{sin2}
\end{equation}%
\ \

Note that in Eq. (\ref{sin2}), DM interactions are combined with only the
first order of $\delta $, and may have much more efficient than $J_{c^{\prime
}i}$ in Eq. (\ref{sin}) since $\delta $ is small around 0.2 \cite{FeGe-1972,FeGe-1977}. This implies
that DM interactions may be the origin of double cone structure.

\subsection{The symmetry analysis of CDW phases}

The high-temperature phase FeGe crystallizes in space group $P6/mmm$, which
has the generators \{3$_{001}^{+}$$|$0\}, \{2$_{001}^{{}}$$|$0\}, \{2$%
_{110}^{{}}$$|$0\} and \{-1$|$0\}, where the left part represents the
rotation and the right part means the lattice translation (here -1 denotes
the inversion symmetry). According to the inversion symmetry, the total
contribution of DM interactions to the energy of double cone magnetic
structure in Eq. (\ref{etot}) is absent, i.e. $\sum_{i,j}D_{ci,j}^{z}=0$,
which is easy to see from the table \ref{relation2}-\ref{relation3}.
Firstly, each kagome layer is still FM in the double cone magnetic state,
thus the in-plane DM interactions are ineffective. For interlayer DM
interactions with an inversion center such as $\mathbf{D}_{c1}$, the
inversion symmetry restricts it to be zero as shown in table \ref{relation2}%
. Meanwhile, for other interlayer DM interactions, the inversion symmetry
combine the equivalent DM interactions in pairs. For example, as shown in
table \ref{relation2}, the $\mathbf{D}_{c2,1}$ and $\mathbf{D}_{c2,7}$\ are
connected by the inversion symmetry, and have opposite values. Therefore,
the summation over equivalent interlayer DM interactions are all zero due to
the inversion symmetry. Note that not only inversion symmetry, but mirror
symmetries such as \{$m_{001}^{{}}$$|$0\}, \{$m_{110}^{{}}$$|$0\}, \{$%
m_{100}^{{}}$$|$0\}, \{$m_{010}^{{}}$$|$0\}, \{$m_{1-10}^{{}}$$|$0\}, \{$%
m_{120}^{{}}$$|$0\} and \{$m_{210}^{{}}$$|$0\} in space group $P6/mmm$,
would also make the DM contribution to the canted magnetic ground state to
be\ zero based on the similar analysis above. Therefore, DM interactions
have no contribution to double cone magnetic structure with the symmetry of
high-temperature phase.

As mentioned in the maintext, since the $2\times 2\times 2$ supercell
structure of CDW phase (compared with the nonmagnetic pristine phase) is
suggested experimentally \cite%
{teng2022discovery,yin2022discovery,2206.12033,2210.06359}, we present the
possible CDW\ phases of kagome FeGe with $2\times 2\times 2$ supercell. The $%
2\times 2\times 2$ supercell without distortion has the symmetry of space
group $P6/mmm$, the non-primitive translation operations $t_{x}$ \{1$|$%
1/2,0,0\}, $t_{y}$ \{1$|$0,1/2,0\}, $t_{z}\ $\{1$|$0,0,1/2\}, and many
symmetry operations from their combinations. As the subgroups compatible
with $2\times 2\times 2$ supercell of pristine FeGe, the structural
distortion of CDW\ phases would break the non-primitive translation
operations $t_{x}$, $t_{y}$ and $t_{z}$, and possibly break other symmetry\
operations as well. Since the point group associated with high-temperature
phase FeGe ($P6/mmm$) is $D_{6h}$, we consider all CDW\ phases whose
associated point group is $D_{6h}$ itself or in maximal subgroups of $D_{6h}$
($D_{2h}$, $D_{6}$, $C_{6h}$, $C_{6v}$, $D_{3d}$, $D_{3h}$). In total we
find 68 different possible CDW phases, and list the corresponding relations
of atomic positions in the high-temperature phase and all types of proposed
CDW phases in table \ref{cdw1}-\ref{cdw5}. Note that the inversion symmetry
and mirror symmetries in parent group $P6/mmm$ would all eliminate the
contribution of DM interactions based on the symmetry analysis. Among these
68 proposed CDW phases, only four distorted structures do not have the
inversion symmetry and mirror symmetries, which can lead to non-zero DM
contribution to double cone spin structure and may explain this magnetic
ground state. They belong to two space groups $P622$ and $P6_{3}22$, and we
list the corresponding Wyckoff positions and the coordinates of the atoms in
the pristine phase and these four CDW phases in table \ref{cdw} of the
maintext.

\begin{table*}[htbp]
\caption{The corresponding Wyckoff positions and the coordinates of the
atoms in the pristine phase and CDW phases with different symmetries. (PART
\uppercase\expandafter{\romannumeral1}). }
\label{cdw1}%
\begin{tabular}{ccccccccccccccc}
\hline
\multicolumn{3}{c|}{Pristine phase(P6/mmm)} & \multicolumn{3}{c|}{
SG191-P6/mmm(type \uppercase\expandafter{\romannumeral1})} &
\multicolumn{3}{c|}{SG191-P6/mmm(type \uppercase\expandafter{\romannumeral2})
} & \multicolumn{3}{c|}{SG194-P6$_{3}$/mmc(type \uppercase%
\expandafter{\romannumeral1})} & \multicolumn{3}{c}{SG194-P6$_{3}$/mmc(type %
\uppercase\expandafter{\romannumeral2})} \\ \hline
& WP & \multicolumn{1}{c|}{Coordinates} &  & WP & \multicolumn{1}{c|}{
Coordinates} &  & WP & \multicolumn{1}{c|}{Coordinates} &  & WP &
\multicolumn{1}{c|}{Coordinates} &  & WP & Coordinates \\ \hline
\multirow{2}{*}{Ge1} & \multirow{2}{*}{1a} & \multicolumn{1}{c|}{%
\multirow{2}{*}{(0, 0, 0)}} & Ge1 & 1a & \multicolumn{1}{c|}{(0, 0, 0)} & Ge1
& 2e & \multicolumn{1}{c|}{(0, 0, z)} & Ge1 & 2a & \multicolumn{1}{c|}{(0,
0, 0)} & Ge1 & 2b & (0, 0, 1/4) \\
&  & \multicolumn{1}{c|}{} & Ge2 & 1b & \multicolumn{1}{c|}{(0, 0, 1/2)} &
Ge2 & 6i & \multicolumn{1}{c|}{(1/2, 0, z)} & Ge2 & 6g & \multicolumn{1}{c|}{
(1/2, 0, 0)} & Ge2 & 6h & (x, 2x, 1/4) \\
\multicolumn{1}{l}{} & \multicolumn{1}{l}{} & \multicolumn{1}{l|}{} & Ge3 &
3f & \multicolumn{1}{c|}{(1/2, 0, 0)} &  &  & \multicolumn{1}{c|}{} &
\multicolumn{1}{l}{} & \multicolumn{1}{l}{} & \multicolumn{1}{l|}{} &
\multicolumn{1}{l}{} & \multicolumn{1}{l}{} & \multicolumn{1}{l}{} \\
\multicolumn{1}{l}{} & \multicolumn{1}{l}{} & \multicolumn{1}{l|}{} & Ge4 &
3g & \multicolumn{1}{c|}{(1/2, 0, 1/2)} &  &  & \multicolumn{1}{c|}{} &
\multicolumn{1}{l}{} & \multicolumn{1}{l}{} & \multicolumn{1}{l|}{} &
\multicolumn{1}{l}{} & \multicolumn{1}{l}{} & \multicolumn{1}{l}{} \\ \hline
Ge2 & 2d & \multicolumn{1}{c|}{(1/3, 2/3, 1/2)} & Ge5 & 4h &
\multicolumn{1}{c|}{(1/3, 2/3, z)} & Ge3 & 2c & \multicolumn{1}{c|}{(1/3,
2/3, 0)} & Ge3 & 2c & \multicolumn{1}{c|}{(1/3, 2/3, 1/4)} & Ge3 & 4f &
(1/3, 2/3, z) \\
&  & \multicolumn{1}{c|}{} & Ge6 & 12o & \multicolumn{1}{c|}{(x, 2x, z)} &
Ge4 & 2d & \multicolumn{1}{c|}{(1/3, 2/3, 1/2)} & Ge4 & 2d &
\multicolumn{1}{c|}{(1/3, 2/3, 1/4)} & Ge4 & 12k & (x, 2x, z) \\
&  & \multicolumn{1}{c|}{} &  &  & \multicolumn{1}{c|}{} & Ge5 & 6l &
\multicolumn{1}{c|}{(x, 2x, 0)} & Ge5 & 6h & \multicolumn{1}{c|}{(x, 2x, 1/4)
} &  &  &  \\
&  & \multicolumn{1}{c|}{} &  &  & \multicolumn{1}{c|}{} & Ge6 & 6m &
\multicolumn{1}{c|}{(x, 2x, 1/2)} & Ge6 & 6h & \multicolumn{1}{c|}{(x, 2x,
1/4)} &  &  &  \\ \hline
\multirow{3}{*}{Fe} & \multirow{3}{*}{3f} & \multicolumn{1}{c|}{%
\multirow{3}{*}{(1/2, 0, 0))}} & Fe1 & 6j & \multicolumn{1}{c|}{(x, 0, 0)} &
Fe1 & 12n & \multicolumn{1}{c|}{(x, 2x, z)} & Fe1 & 12k &
\multicolumn{1}{c|}{(x, 0, 0)} & Fe1 & 6h & (x, 2x, 1/4) \\
&  & \multicolumn{1}{c|}{} & Fe2 & 6k & \multicolumn{1}{c|}{(x, 0, 1/2)} &
Fe2 & 12o & \multicolumn{1}{c|}{(x, 0, z)} & Fe2 & 12k & \multicolumn{1}{c|}{
(x, 2x, z)} & Fe2 & 6h & (x, 2x, 1/4) \\
&  & \multicolumn{1}{c|}{} & Fe3 & 6l & \multicolumn{1}{c|}{(x, 2x, 0)} &  &
& \multicolumn{1}{c|}{} &  &  & \multicolumn{1}{c|}{} & Fe3 & 12j & (x, y,
1/4) \\
\multicolumn{1}{l}{} & \multicolumn{1}{l}{} & \multicolumn{1}{l|}{} & Fe4 &
6m & \multicolumn{1}{c|}{(x, 2x, 1/2)} &  &  & \multicolumn{1}{c|}{} &
\multicolumn{1}{l}{} & \multicolumn{1}{l}{} & \multicolumn{1}{l|}{} &
\multicolumn{1}{l}{} & \multicolumn{1}{l}{} & \multicolumn{1}{l}{} \\ \hline
&  &  &  &  &  &  &  &  &  &  &  &  &  &  \\ \hline
\multicolumn{3}{c|}{Pristine phase(P6/mmm)} & \multicolumn{3}{c|}{SG193-P6$%
_{3}$/mcm(type \uppercase\expandafter{\romannumeral1})} &
\multicolumn{3}{c|}{SG193-P6$_{3}$/mcm(type \uppercase\expandafter{%
\romannumeral2})} & \multicolumn{3}{c|}{SG192-P6/mcc(type \uppercase%
\expandafter{\romannumeral1})} & \multicolumn{3}{c}{SG192-P6/mcc(type %
\uppercase\expandafter{\romannumeral2})} \\ \hline
& WP & \multicolumn{1}{c|}{Coordinates} &  & WP & \multicolumn{1}{c|}{
Coordinates} &  & WP & \multicolumn{1}{c|}{Coordinates} &  &  &
\multicolumn{1}{c|}{} &  & WP & Coordinates \\ \hline
\multirow{2}{*}{Ge1} & \multirow{2}{*}{1a} & \multicolumn{1}{c|}{%
\multirow{2}{*}{(0, 0, 0)}} & Ge1 & 2b & \multicolumn{1}{c|}{(0, 0, 0)} & Ge1
& 2a & \multicolumn{1}{c|}{(0, 0, 1/4)} & Ge1 & 2b & \multicolumn{1}{c|}{(0,
0, 0)} & Ge1 & 2b & (0, 0, 1/4) \\
&  & \multicolumn{1}{c|}{} & Ge2 & 6f & \multicolumn{1}{c|}{(1/2, 0, 0)} &
Ge2 & 6g & \multicolumn{1}{c|}{(x, 0, 1/4)} & Ge2 & 6g & \multicolumn{1}{c|}{
(1/2, 0, 0)} & Ge2 & 6f & (1/2, 0, 1/4) \\ \hline
\multirow{2}{*}{Ge2} & \multirow{2}{*}{2d} & \multicolumn{1}{c|}{%
\multirow{2}{*}{(1/3, 2/3, 1/2)}} & Ge3 & 4c & \multicolumn{1}{c|}{(1/3,
2/3, 1/4)} & Ge3 & 4d & \multicolumn{1}{c|}{(1/3, 2/3, 0)} & Ge3 & 4c &
\multicolumn{1}{c|}{(1/3, 2/3, 1/4)} & Ge3 & 4d & (1/3, 2/3, z) \\
&  & \multicolumn{1}{c|}{} & Ge4 & 12j & \multicolumn{1}{c|}{(x, y, 1/4)} &
Ge4 & 12i & \multicolumn{1}{c|}{(x, 2x, 0)} & Ge4 & 12k &
\multicolumn{1}{c|}{(x, 2x, 1/4)} & Ge4 & 12l & (x, y, 0) \\ \hline
\multirow{3}{*}{Fe} & \multirow{3}{*}{3f} & \multicolumn{1}{c|}{%
\multirow{3}{*}{(1/2, 0, 0))}} & Fe1 & 12i & \multicolumn{1}{c|}{(x, 0, z)}
& Fe1 & 6g & \multicolumn{1}{c|}{(x, 0, 1/4)} & Fe1 & 12l &
\multicolumn{1}{c|}{(x, y, 0)} & Fe1 & 12j & (x, 0, 1/4) \\
&  & \multicolumn{1}{c|}{} & Fe2 & 12k & \multicolumn{1}{c|}{(x 2x ,0)} & Fe2
& 6g & \multicolumn{1}{c|}{(x, 0, 1/4)} & Fe2 & 12l & \multicolumn{1}{c|}{
(x, y, 0)} & Fe2 & 12k & (x, 2x, 1/4) \\
&  & \multicolumn{1}{c|}{} &  &  & \multicolumn{1}{c|}{} & Fe3 & 12j &
\multicolumn{1}{c|}{(x, y, 1/4)} &  &  & \multicolumn{1}{c|}{} &  &  &  \\
\hline
&  &  &  &  &  &  &  &  &  &  &  &  &  &  \\ \hline
\multicolumn{3}{c|}{Pristine phase(P6/mmm)} & \multicolumn{3}{c|}{SG190-P$%
\overline{6}$2c(type \uppercase\expandafter{\romannumeral1})} &
\multicolumn{3}{c|}{SG190-P$\overline{6}$2c(type \uppercase%
\expandafter{\romannumeral2})} & \multicolumn{3}{c|}{SG189-P$\overline{6}$%
2m(type \uppercase\expandafter{\romannumeral1})} & \multicolumn{3}{c}{SG189-P%
$\overline{6}$2m(type \uppercase\expandafter{\romannumeral2})} \\ \hline
& WP & \multicolumn{1}{c|}{Coordinates} & \multicolumn{1}{c}{} &
\multicolumn{1}{c}{WP} & \multicolumn{1}{c|}{Coordinates} &
\multicolumn{1}{c}{} & \multicolumn{1}{c}{WP} & \multicolumn{1}{c|}{
Coordinates} & \multicolumn{1}{c}{} & \multicolumn{1}{c}{WP} &
\multicolumn{1}{c|}{Coordinates} & \multicolumn{1}{c}{} & \multicolumn{1}{c}{
WP} & \multicolumn{1}{c}{Coordinates} \\ \hline
\multirow{4}{*}{Ge1} & \multirow{4}{*}{1a} & \multicolumn{1}{c|}{%
\multirow{4}{*}{(0, 0, 0)}} & Ge1 & 2a & \multicolumn{1}{l|}{(0, 0, 0)} & Ge1
& 2b & \multicolumn{1}{l|}{(0, 0, 1/4)} & Ge1 & 1a & \multicolumn{1}{l|}{(0,
0, 0)} & Ge1 & 2e & (0, 0, z) \\
&  & \multicolumn{1}{c|}{} & Ge2 & 6g & \multicolumn{1}{l|}{(x, 0, 0)} & Ge2
& 6h & \multicolumn{1}{l|}{(x, y, 1/4)} & Ge2 & 1b & \multicolumn{1}{l|}{(0,
0, 1/2)} & Ge2 & 6i & (x, 0, z) \\
&  & \multicolumn{1}{c|}{} &  &  & \multicolumn{1}{l|}{} &  &  &
\multicolumn{1}{l|}{} & Ge3 & 3f & \multicolumn{1}{l|}{(x, 0, 0)} &  &  &
\\
&  & \multicolumn{1}{c|}{} &  &  & \multicolumn{1}{l|}{} &  &  &
\multicolumn{1}{l|}{} & Ge4 & 3g & \multicolumn{1}{l|}{(x, 0, 1/2)} &  &  &
\\ \hline
\multirow{4}{*}{Ge2} & \multirow{4}{*}{2d} & \multicolumn{1}{c|}{%
\multirow{4}{*}{(1/3, 2/3, 1/2)}} & Ge3 & 2c & \multicolumn{1}{l|}{(1/3,
2/3, 1/4)} & Ge3 & 4f & \multicolumn{1}{l|}{(1/3, 2/3, z)} & Ge5 & 4h &
\multicolumn{1}{l|}{(1/3, 2/3, z)} & Ge3 & 2c & (1/3, 2/3, 0) \\
&  & \multicolumn{1}{c|}{} & Ge4 & 2d & \multicolumn{1}{l|}{(1/3, 2/3, 3/4)}
& Ge4 & 12i & \multicolumn{1}{l|}{(x, y, z)} & Ge6 & 12l &
\multicolumn{1}{l|}{(x, y, z)} & Ge4 & 2d & (1/3, 2/3, 1/2) \\
&  & \multicolumn{1}{c|}{} & Ge5 & 6h & \multicolumn{1}{l|}{(x, y, 1/4)} &
&  & \multicolumn{1}{l|}{} &  &  & \multicolumn{1}{l|}{} & Ge5 & 6j & (x, y,
0) \\
&  & \multicolumn{1}{c|}{} & Ge6 & 6h & \multicolumn{1}{l|}{(x, y, 1/4)} &
&  & \multicolumn{1}{l|}{} &  &  & \multicolumn{1}{l|}{} & Ge6 & 6k & (x, y,
1/2) \\ \hline
\multirow{6}{*}{Fe} & \multirow{6}{*}{3f} & \multicolumn{1}{c|}{%
\multirow{6}{*}{(1/2, 0, 0))}} & Fe1 & 6g & \multicolumn{1}{l|}{(x, 0, 0)} &
Fe1 & 6h & \multicolumn{1}{l|}{(x, y, 1/4)} & Fe1 & 3f & \multicolumn{1}{l|}{
(x, 0, 0)} & Fe1 & 6i & (x, 0, z) \\
&  & \multicolumn{1}{c|}{} & Fe2 & 6g & \multicolumn{1}{l|}{(x, 0, 0)} & Fe2
& 6h & \multicolumn{1}{l|}{(x, y, 1/4)} & Fe2 & 3f & \multicolumn{1}{l|}{(x,
0, 0)} & Fe2 & 6i & (x, 0, z) \\
&  & \multicolumn{1}{c|}{} & Fe3 & 12i & \multicolumn{1}{l|}{(x, y, z)} & Fe3
& 6h & \multicolumn{1}{l|}{(x, y, 1/4)} & Fe3 & 3g & \multicolumn{1}{l|}{(x,
0, 1/2)} & Fe3 & 12l & (x, y, z) \\
&  & \multicolumn{1}{c|}{} &  &  & \multicolumn{1}{l|}{} & Fe4 & 6h &
\multicolumn{1}{l|}{(x, y, 1/4)} & Fe4 & 3g & \multicolumn{1}{l|}{(x, 0, 1/2)
} &  &  &  \\
&  & \multicolumn{1}{c|}{} &  &  & \multicolumn{1}{l|}{} &  &  &
\multicolumn{1}{l|}{} & Fe5 & 6j & \multicolumn{1}{l|}{(x, y, 0)} &  &  &
\\
&  & \multicolumn{1}{c|}{} &  &  & \multicolumn{1}{l|}{} &  &  &
\multicolumn{1}{l|}{} & Fe6 & 6k & \multicolumn{1}{l|}{(x, y, 1/2)} &  &  &
\\ \hline
\multicolumn{1}{l}{} & \multicolumn{1}{l}{} & \multicolumn{1}{l}{} &  &  &
&  &  &  &  &  &  &  &  &  \\ \hline
\multicolumn{3}{c|}{Pristine phase(P6/mmm)} & \multicolumn{3}{c|}{SG188-P$%
\overline{6}$c2(type \uppercase\expandafter{\romannumeral1})} &
\multicolumn{3}{c|}{SG188-P$\overline{6}$c2(type \uppercase%
\expandafter{\romannumeral2})} & \multicolumn{3}{c|}{SG187-P$\overline{6}$%
m2(type \uppercase\expandafter{\romannumeral1})} & \multicolumn{3}{c}{SG187-P%
$\overline{6}$m2(type \uppercase\expandafter{\romannumeral2})} \\ \hline
& WP & \multicolumn{1}{c|}{Coordinates} & \multicolumn{1}{c}{} &
\multicolumn{1}{c}{WP} & \multicolumn{1}{c|}{Coordinates} &
\multicolumn{1}{c}{} & \multicolumn{1}{c}{WP} & \multicolumn{1}{c|}{
Coordinates} & \multicolumn{1}{c}{} & \multicolumn{1}{c}{WP} &
\multicolumn{1}{c|}{Coordinates} & \multicolumn{1}{c}{} & \multicolumn{1}{c}{
WP} & \multicolumn{1}{c}{Coordinates} \\ \hline
\multicolumn{1}{l}{\multirow{4}{*}{Ge1}} & \multicolumn{1}{l}{%
\multirow{4}{*}{1a}} & \multicolumn{1}{l|}{\multirow{4}{*}{(0, 0, 0)}} & Ge1
& 2a & \multicolumn{1}{l|}{(0, 0, 0)} & Ge1 & 2d & \multicolumn{1}{l|}{(1/3,
2/3, 1/4)} & Ge1 & 1a & \multicolumn{1}{l|}{(0, 0, 0)} & Ge1 & 2h & (1/3,
2/3, z) \\
\multicolumn{1}{l}{} & \multicolumn{1}{l}{} & \multicolumn{1}{l|}{} & Ge2 &
6j & \multicolumn{1}{l|}{(x, 2x, 0)} & Ge2 & 6k & \multicolumn{1}{l|}{(x, y,
1/4)} & Ge2 & 1b & \multicolumn{1}{l|}{(0, 0, 1/2)} & Ge2 & 6n & (x, 2x, z)
\\
\multicolumn{1}{l}{} & \multicolumn{1}{l}{} & \multicolumn{1}{l|}{} &  &  &
\multicolumn{1}{l|}{} &  &  & \multicolumn{1}{l|}{} & Ge3 & 3j &
\multicolumn{1}{l|}{(x, 2x, 0)} &  &  &  \\
\multicolumn{1}{l}{} & \multicolumn{1}{l}{} & \multicolumn{1}{l|}{} &  &  &
\multicolumn{1}{l|}{} &  &  & \multicolumn{1}{l|}{} & Ge4 & 3k &
\multicolumn{1}{l|}{(x, 2x, 1/2)} &  &  &  \\ \hline
\multicolumn{1}{l}{\multirow{8}{*}{Ge2}} & \multicolumn{1}{l}{%
\multirow{8}{*}{2d}} & \multicolumn{1}{l|}{\multirow{8}{*}{(1/3, 2/3, 1/2)}}
& Ge3 & 2d & \multicolumn{1}{l|}{(2/3, 1/3, 1/4)} & Ge3 & 2a &
\multicolumn{1}{l|}{(0, 0, 0)} & Ge5 & 2i & \multicolumn{1}{l|}{(2/3, 1/3, z)
} & Ge3 & 1a & (0, 0, 0) \\
\multicolumn{1}{l}{} & \multicolumn{1}{l}{} & \multicolumn{1}{l|}{} & Ge4 &
2f & \multicolumn{1}{l|}{(1/3, 2/3, 1/4)} & Ge4 & 2e & \multicolumn{1}{l|}{
(2/3, 1/3, 0)} & Ge6 & 2h & \multicolumn{1}{l|}{(1/3, 2/3, z)} & Ge4 & 1b &
(0, 0, 1/2) \\
\multicolumn{1}{l}{} & \multicolumn{1}{l}{} & \multicolumn{1}{l|}{} & Ge5 &
6k & \multicolumn{1}{l|}{(x, y, 1/4)} & Ge5 & 6j & \multicolumn{1}{l|}{(x,
2x, 0)} & Ge7 & 6n & \multicolumn{1}{l|}{(x, 2x, z)} & Ge5 & 1e & (2/3, 1/3,
0) \\
\multicolumn{1}{l}{} & \multicolumn{1}{l}{} & \multicolumn{1}{l|}{} & Ge6 &
6k & \multicolumn{1}{l|}{(x, y, 1/4)} & Ge6 & 6j & \multicolumn{1}{l|}{(x,
2x, 1/2)} & Ge8 & 6n & \multicolumn{1}{l|}{(x, 2x, z)} & Ge6 & 1f & (2/3,
1/3, 1/2) \\
\multicolumn{1}{l}{} & \multicolumn{1}{l}{} & \multicolumn{1}{l|}{} &  &  &
\multicolumn{1}{l|}{} &  &  & \multicolumn{1}{l|}{} &  &  &
\multicolumn{1}{l|}{} & Ge7 & 3j & (x, 2x, 0) \\
\multicolumn{1}{l}{} & \multicolumn{1}{l}{} & \multicolumn{1}{l|}{} &  &  &
\multicolumn{1}{l|}{} &  &  & \multicolumn{1}{l|}{} &  &  &
\multicolumn{1}{l|}{} & Ge8 & 3j & (x, 2x, 0) \\
\multicolumn{1}{l}{} & \multicolumn{1}{l}{} & \multicolumn{1}{l|}{} &  &  &
\multicolumn{1}{l|}{} &  &  & \multicolumn{1}{l|}{} &  &  &
\multicolumn{1}{l|}{} & Ge9 & 3k & (x, 2x, 1/2) \\
\multicolumn{1}{l}{} & \multicolumn{1}{l}{} & \multicolumn{1}{l|}{} &  &  &
\multicolumn{1}{l|}{} &  &  & \multicolumn{1}{l|}{} &  &  &
\multicolumn{1}{l|}{} & Ge10 & 3k & (x, 2x, 1/2) \\ \hline
\multicolumn{1}{l}{\multirow{6}{*}{Fe}} & \multicolumn{1}{l}{%
\multirow{6}{*}{3f}} & \multicolumn{1}{l|}{\multirow{6}{*}{(1/2, 0, 0)}} &
Fe1 & 6j & \multicolumn{1}{l|}{(x, 2x, 0)} & Fe1 & 6k & \multicolumn{1}{l|}{
(x, y, 1/4)} & Fe1 & 3j & \multicolumn{1}{l|}{(x, 2x, 0)} & Fe1 & 6n & (x,
2x ,z) \\
\multicolumn{1}{l}{} & \multicolumn{1}{l}{} & \multicolumn{1}{l|}{} & Fe2 &
6j & \multicolumn{1}{l|}{(x, 2x, 0)} & Fe2 & 6k & \multicolumn{1}{l|}{(x, y,
1/4)} & Fe2 & 3j & \multicolumn{1}{l|}{(x, 2x, 0)} & Fe2 & 6n & (x, 2x ,z)
\\
\multicolumn{1}{l}{} & \multicolumn{1}{l}{} & \multicolumn{1}{l|}{} & Fe3 &
12l & \multicolumn{1}{l|}{(x, y, z)} & Fe3 & 6k & \multicolumn{1}{l|}{(x, y,
1/4)} & Fe3 & 3k & \multicolumn{1}{l|}{(x, 2x, 1/2)} & Fe3 & 12o & (x, y, z)
\\
\multicolumn{1}{l}{} & \multicolumn{1}{l}{} & \multicolumn{1}{l|}{} &  &  &
\multicolumn{1}{l|}{} & Fe4 & 6k & \multicolumn{1}{l|}{(x, y, 1/4)} & Fe4 &
3k & \multicolumn{1}{l|}{(x, 2x, 1/2)} &  &  &  \\
\multicolumn{1}{l}{} & \multicolumn{1}{l}{} & \multicolumn{1}{l|}{} &  &  &
\multicolumn{1}{l|}{} &  &  & \multicolumn{1}{l|}{} & Fe5 & 6l &
\multicolumn{1}{l|}{(x, y, 0)} &  &  &  \\
\multicolumn{1}{l}{} & \multicolumn{1}{l}{} & \multicolumn{1}{l|}{} &  &  &
\multicolumn{1}{l|}{} &  &  & \multicolumn{1}{l|}{} & Fe6 & 6m &
\multicolumn{1}{l|}{(x, y, 1/2)} &  &  &  \\ \hline
\end{tabular}%
\end{table*}

\begin{table*}[htbp]
\caption{The corresponding Wyckoff positions and the coordinates of the
atoms in the pristine phase and CDW phases with different symmetries. (PART
\uppercase\expandafter{\romannumeral2}).}
\label{cdw2}%
\begin{tabular}{ccccccccccccccc}
\hline
\multicolumn{3}{c|}{Pristine phase(P6/mmm)} & \multicolumn{3}{c|}{SG186-P$%
\overline{6}_{3}$mc(type \uppercase\expandafter{\romannumeral1})} &
\multicolumn{3}{c|}{SG185-P$\overline{6}_{3}$cm(type \uppercase%
\expandafter{\romannumeral1})} & \multicolumn{3}{c|}{SG184-P6cc(type %
\uppercase\expandafter{\romannumeral1})} & \multicolumn{3}{c}{
SG183-P6mm(type \uppercase\expandafter{\romannumeral1})} \\ \hline
& WP & \multicolumn{1}{c|}{Coordinates} &  & WP & \multicolumn{1}{c|}{
Coordinates} &  & WP & \multicolumn{1}{c|}{Coordinates} &  & WP &
\multicolumn{1}{c|}{Coordinates} &  & WP & Coordinates \\ \hline
\multirow{4}{*}{Ge1} & \multirow{4}{*}{1a} & \multicolumn{1}{c|}{%
\multirow{4}{*}{(0, 0, 0)}} & Ge1 & 2a & \multicolumn{1}{c|}{(0, 0, z)} & Ge1
& 2a & \multicolumn{1}{c|}{(0, 0, z)} & Ge1 & 2a & \multicolumn{1}{c|}{(0,
0, z)} & Ge1 & 1a & (0, 0, z) \\
&  & \multicolumn{1}{c|}{} & Ge2 & 6c & \multicolumn{1}{c|}{(x, 0, z)} & Ge2
& 6c & \multicolumn{1}{c|}{(x, 2x, z)} & Ge2 & 6c & \multicolumn{1}{c|}{
(1/2, 0, z)} & Ge2 & 1a & (0, 0, z) \\
&  & \multicolumn{1}{c|}{} &  &  & \multicolumn{1}{c|}{} &  &  &
\multicolumn{1}{c|}{} &  &  & \multicolumn{1}{c|}{} & Ge3 & 3c & (1/2, 0, z)
\\
&  & \multicolumn{1}{c|}{} &  &  & \multicolumn{1}{c|}{} &  &  &
\multicolumn{1}{c|}{} &  &  & \multicolumn{1}{c|}{} & Ge4 & 3c & (1/2, 0, z)
\\ \hline
\multirow{4}{*}{Ge2} & \multirow{4}{*}{2d} & \multicolumn{1}{c|}{%
\multirow{4}{*}{(1/3, 2/3, 1/2)}} & Ge3 & 4b & \multicolumn{1}{c|}{(1/3,
2/3, z)} & Ge3 & 2b & \multicolumn{1}{c|}{(1/3, 2/3, z)} & Ge3 & 4b &
\multicolumn{1}{c|}{(1/3, 2/3, z)} & Ge5 & 2b & (1/3, 2/3, z) \\
&  & \multicolumn{1}{c|}{} & Ge4 & 12d & \multicolumn{1}{c|}{(x, y, z)} & Ge4
& 2b & \multicolumn{1}{c|}{(1/3, 2/3, z)} & Ge4 & 12d & \multicolumn{1}{c|}{
(x, y, z)} & Ge6 & 2b & (1/3, 2/3, z) \\
&  & \multicolumn{1}{c|}{} &  &  & \multicolumn{1}{c|}{} & Ge5 & 6c &
\multicolumn{1}{c|}{(x, 2x, z)} &  &  & \multicolumn{1}{c|}{} & Ge7 & 6e &
(x, 2x, z) \\
&  & \multicolumn{1}{c|}{} &  &  & \multicolumn{1}{c|}{} & Ge6 & 6c &
\multicolumn{1}{c|}{(x, 2x, z)} &  &  & \multicolumn{1}{c|}{} & Ge8 & 6e &
(x, 2x, z) \\ \hline
\multirow{4}{*}{Fe} & \multirow{4}{*}{3f} & \multicolumn{1}{c|}{%
\multirow{4}{*}{(1/2, 0, 0)}} & Fe1 & 6c & \multicolumn{1}{c|}{(x, 0, z)} &
Fe1 & 6c & \multicolumn{1}{c|}{(x, 2x, z)} & Fe1 & 12d & \multicolumn{1}{c|}{
(x, y, z)} & Fe1 & 6d & (x, 0, z) \\
&  & \multicolumn{1}{c|}{} & Fe2 & 6c & \multicolumn{1}{c|}{(x, 0, z)} & Fe2
& 6c & \multicolumn{1}{c|}{(x, 2x, z)} & Fe2 & 12d & \multicolumn{1}{c|}{(x,
y, z)} & Fe2 & 6d & (x, 0, z) \\
&  & \multicolumn{1}{c|}{} & Fe3 & 12d & \multicolumn{1}{c|}{(x, y, z)} & Fe3
& 12d & \multicolumn{1}{c|}{(x, y, z)} &  &  & \multicolumn{1}{c|}{} & Fe3 &
6d & (x, 2x, z) \\
&  & \multicolumn{1}{c|}{} &  &  & \multicolumn{1}{c|}{} &  &  &
\multicolumn{1}{c|}{} &  &  & \multicolumn{1}{c|}{} & Fe4 & 6d & (x, 2x, z)
\\ \hline
&  &  &  &  &  &  &  &  &  &  &  &  &  &  \\ \hline
\multicolumn{3}{c|}{Pristine phase(P6/mmm)} & \multicolumn{3}{c|}{SG182-P6$%
_{3}$22(type \uppercase\expandafter{\romannumeral1})} & \multicolumn{3}{c|}{
SG182-P6$_{3}$22(type \uppercase\expandafter{\romannumeral2})} &
\multicolumn{3}{c|}{SG177-P622(type \uppercase\expandafter{\romannumeral1})}
& \multicolumn{3}{c}{SG177-P622(type \uppercase\expandafter{\romannumeral2})}
\\ \hline
& WP & \multicolumn{1}{c|}{Coordinates} &  & WP & \multicolumn{1}{c|}{
Coordinates} &  & WP & \multicolumn{1}{c|}{Coordinates} &  & WP &
\multicolumn{1}{c|}{Coordinates} &  & WP & Coordinates \\ \hline
\multirow{4}{*}{Ge1} & \multirow{4}{*}{1a} & \multicolumn{1}{c|}{%
\multirow{4}{*}{(0, 0, 0)}} & Ge1 & 2a & \multicolumn{1}{c|}{(0, 0, 0)} & Ge1
& 2b & \multicolumn{1}{c|}{(0, 0, 1/4)} & Ge1 & 1a & \multicolumn{1}{c|}{(0,
0, 0)} & Ge1 & 2e & (0, 0, z) \\
&  & \multicolumn{1}{c|}{} & Ge2 & 6g & \multicolumn{1}{c|}{(x, 0, 0)} & Ge2
& 6h & \multicolumn{1}{c|}{(x, 2x, 1/4)} & Ge2 & 1b & \multicolumn{1}{c|}{
(0, 0, 1/2)} & Ge2 & 6i & (1/2, 0, z) \\
&  & \multicolumn{1}{c|}{} &  &  & \multicolumn{1}{c|}{} &  &  &
\multicolumn{1}{c|}{} & Ge3 & 3f & \multicolumn{1}{c|}{(0, 1/2, 0)} &  &  &
\\
&  & \multicolumn{1}{c|}{} &  &  & \multicolumn{1}{c|}{} &  &  &
\multicolumn{1}{c|}{} & Ge4 & 3g & \multicolumn{1}{c|}{(0, 1/2, 1/2)} &  &
&  \\ \hline
\multirow{4}{*}{Ge2} & \multirow{4}{*}{2d} & \multicolumn{1}{c|}{%
\multirow{4}{*}{(1/3, 2/3, 1/2)}} & Ge3 & 2c & \multicolumn{1}{c|}{(1/3,
2/3, 1/4)} & Ge3 & 4f & \multicolumn{1}{c|}{(1/3, 2/3, z)} & Ge5 & 4h &
\multicolumn{1}{c|}{(1/3, 2/3, z)} & Ge3 & 2c & (1/3, 2/3, 0) \\
&  & \multicolumn{1}{c|}{} & Ge4 & 2d & \multicolumn{1}{c|}{(1/3, 2/3, 3/4)}
& Ge4 & 12i & \multicolumn{1}{c|}{(x, y, z)} & Ge6 & 12n &
\multicolumn{1}{c|}{(x, y, z)} & Ge4 & 2d & (1/3, 2/3,1/2) \\
&  & \multicolumn{1}{c|}{} & Ge5 & 6h & \multicolumn{1}{c|}{(x, 2x, 1/4)} &
&  & \multicolumn{1}{c|}{} &  &  & \multicolumn{1}{c|}{} & Ge5 & 6l & (x,
2x, 0) \\
&  & \multicolumn{1}{c|}{} & Ge6 & 6h & \multicolumn{1}{c|}{(x, 2x, 1/4)} &
&  & \multicolumn{1}{c|}{} &  &  & \multicolumn{1}{c|}{} & Ge6 & 6m & (x,
2x, 1/2) \\ \hline
\multirow{4}{*}{Fe} & \multirow{4}{*}{3f} & \multicolumn{1}{c|}{%
\multirow{4}{*}{(1/2, 0, 0)}} & Fe1 & 6g & \multicolumn{1}{c|}{(x, 0, 0)} &
Fe1 & 6h & \multicolumn{1}{c|}{(x, 2x, 1/4)} & Fe1 & 6j &
\multicolumn{1}{c|}{(x, 0, 0)} & Fe1 & 12n & (x, y, z) \\
&  & \multicolumn{1}{c|}{} & Fe2 & 6g & \multicolumn{1}{c|}{(x, 0, 0)} & Fe2
& 6h & \multicolumn{1}{c|}{(x, 2x, 1/4)} & Fe2 & 6k & \multicolumn{1}{c|}{
(x, 0, 1/2)} & Fe1 & 12n & (x, y, z) \\
&  & \multicolumn{1}{c|}{} & Fe3 & 12i & \multicolumn{1}{c|}{(x, y, z)} & Fe3
& 12i & \multicolumn{1}{c|}{(x, y, z)} & Fe3 & 6l & \multicolumn{1}{c|}{(x,
2x, 0)} &  &  &  \\
&  & \multicolumn{1}{c|}{} &  &  & \multicolumn{1}{c|}{} &  &  &
\multicolumn{1}{c|}{} & Fe4 & 6m & \multicolumn{1}{c|}{(x, 2x, 1/2)} &  &  &
\\ \hline
&  &  &  &  &  &  &  &  &  &  &  &  &  &  \\ \hline
\multicolumn{3}{c|}{Pristine phase(P6/mmm)} & \multicolumn{3}{c|}{SG176-P6$%
_{3}$/m(type \uppercase\expandafter{\romannumeral1})} & \multicolumn{3}{c|}{
SG176-P6$_{3}$/m(type \uppercase\expandafter{\romannumeral2})} &
\multicolumn{3}{c|}{SG175-P6/m(type \uppercase\expandafter{\romannumeral1})}
& \multicolumn{3}{c}{SG175-P6/m(type \uppercase\expandafter{\romannumeral2})}
\\ \hline
& WP & \multicolumn{1}{c|}{Coordinates} &  & WP & \multicolumn{1}{c|}{
Coordinates} &  & WP & \multicolumn{1}{c|}{Coordinates} &  & WP &
\multicolumn{1}{c|}{Coordinates} &  & WP & Coordinates \\ \hline
\multirow{4}{*}{Ge1} & \multirow{4}{*}{1a} & \multicolumn{1}{c|}{%
\multirow{4}{*}{(0, 0, 0)}} & Ge1 & 2b & \multicolumn{1}{c|}{(0, 0, 0)} & Ge1
& 2a & \multicolumn{1}{c|}{(0, 0, 1/4)} & Ge1 & 1a & \multicolumn{1}{c|}{(0,
0, 0)} & Ge1 & 2e & (0, 1/2, z) \\
&  & \multicolumn{1}{c|}{} & Ge2 & 6g & \multicolumn{1}{c|}{(1/2, 0, 0)} &
Ge2 & 6h & \multicolumn{1}{c|}{(x, y, 1/4)} & Ge2 & 1b & \multicolumn{1}{c|}{
(0, 0, 1/2)} & Ge2 & 6i & (0, 0, z) \\
&  & \multicolumn{1}{c|}{} &  &  & \multicolumn{1}{c|}{} &  &  &
\multicolumn{1}{c|}{} & Ge3 & 3f & \multicolumn{1}{c|}{(1/2, 0, 0)} &  &  &
\\
&  & \multicolumn{1}{c|}{} &  &  & \multicolumn{1}{c|}{} &  &  &
\multicolumn{1}{c|}{} & Ge4 & 3g & \multicolumn{1}{c|}{(1/2, 0, 1/2)} &  &
&  \\ \hline
\multirow{4}{*}{Ge2} & \multirow{4}{*}{2d} & \multicolumn{1}{c|}{%
\multirow{4}{*}{(1/3, 2/3, 1/2)}} & Ge3 & 2c & \multicolumn{1}{c|}{(1/3,
2/3, 1/4)} & Ge3 & 4f & \multicolumn{1}{c|}{(1/3, 2/3, z)} & Ge5 & 4h &
\multicolumn{1}{c|}{(1/3, 2/3, z)} & Ge3 & 2c & (1/3, 2/3, 0) \\
&  & \multicolumn{1}{c|}{} & Ge4 & 2d & \multicolumn{1}{c|}{(1/3, 2/3, 3/4)}
& Ge4 & 12i & \multicolumn{1}{c|}{(x, y, z)} & Ge6 & 12l &
\multicolumn{1}{c|}{(x, y, z)} & Ge4 & 2d & (1/3, 2/3, 1/2) \\
&  & \multicolumn{1}{c|}{} & Ge5 & 6h & \multicolumn{1}{c|}{(x, y, 1/4)} &
&  & \multicolumn{1}{c|}{} &  &  & \multicolumn{1}{c|}{} & Ge5 & 6j & (x, y,
0) \\
&  & \multicolumn{1}{c|}{} & Ge6 & 6h & \multicolumn{1}{c|}{(x, y, 1/4)} &
&  & \multicolumn{1}{c|}{} &  &  & \multicolumn{1}{c|}{} & Ge6 & 6k & (x, y,
1/2) \\ \hline
\multirow{4}{*}{Fe} & \multirow{4}{*}{3f} & \multicolumn{1}{c|}{%
\multirow{4}{*}{(1/2, 0, 0)}} & Fe1 & 12i & \multicolumn{1}{c|}{(x, y, z)} &
Fe1 & 6h & \multicolumn{1}{c|}{(x, y, 1/4)} & Fe1 & 6j & \multicolumn{1}{c|}{
(x, y, 0)} & Fe1 & 12l & (x, y, z) \\
&  & \multicolumn{1}{c|}{} & Fe2 & 12i & \multicolumn{1}{c|}{(x, y, z)} & Fe2
& 6h & \multicolumn{1}{c|}{(x, y, 1/4)} & Fe2 & 6j & \multicolumn{1}{c|}{(x,
y, 0)} & Fe2 & 12l & (x, y, z) \\
&  & \multicolumn{1}{c|}{} &  &  & \multicolumn{1}{c|}{} & Fe3 & 6h &
\multicolumn{1}{c|}{(x, y, 1/4)} & Fe3 & 6k & \multicolumn{1}{c|}{(x, y, 1/2)
} &  &  &  \\
&  & \multicolumn{1}{c|}{} &  &  & \multicolumn{1}{c|}{} & Fe4 & 6h &
\multicolumn{1}{c|}{(x, y, 1/4)} & Fe4 & 6k & \multicolumn{1}{c|}{(x, y, 1/2)
} &  &  &  \\ \hline
&  &  &  &  &  &  &  &  &  &  &  &  &  &  \\ \hline
\multicolumn{3}{c|}{Pristine phase(P6/mmm)} & \multicolumn{3}{c|}{SG165-P$%
\overline{3}$c1(type \uppercase\expandafter{\romannumeral1})} &
\multicolumn{3}{c|}{SG165-P$\overline{3}$c1(type \uppercase%
\expandafter{\romannumeral2})} & \multicolumn{3}{c|}{SG164-P$\overline{3}$%
m1(type \uppercase\expandafter{\romannumeral1})} & \multicolumn{3}{c}{SG164-P%
$\overline{3}$m1(type \uppercase\expandafter{\romannumeral2})} \\ \hline
& WP & \multicolumn{1}{c|}{Coordinates} &  & WP & \multicolumn{1}{c|}{
Coordinates} &  & WP & \multicolumn{1}{c|}{Coordinates} &  & WP &
\multicolumn{1}{c|}{Coordinates} &  & WP & Coordinates \\ \hline
\multirow{4}{*}{Ge1} & \multirow{4}{*}{1a} & \multicolumn{1}{c|}{%
\multirow{4}{*}{(0, 0, 0)}} & Ge1 & 2b & \multicolumn{1}{c|}{(0, 0, 0)} & Ge1
& 2a & \multicolumn{1}{c|}{(0, 0, 1/4)} & Ge1 & 1a & \multicolumn{1}{c|}{(0,
0, 0)} & Ge1 & 2c & (0, 0, z) \\
&  & \multicolumn{1}{c|}{} & Ge2 & 6e & \multicolumn{1}{c|}{(1/2, 0, 0)} &
Ge2 & 6f & \multicolumn{1}{c|}{(x, 0, 1/4)} & Ge2 & 1b & \multicolumn{1}{c|}{
(0, 0, 1/2)} & Ge2 & 6i & (x, 2x z) \\
&  & \multicolumn{1}{c|}{} &  &  & \multicolumn{1}{c|}{} &  &  &
\multicolumn{1}{c|}{} & Ge3 & 3e & \multicolumn{1}{c|}{(0, 1/2, 0)} &  &  &
\\
&  & \multicolumn{1}{c|}{} &  &  & \multicolumn{1}{c|}{} &  &  &
\multicolumn{1}{c|}{} & Ge4 & 3f & \multicolumn{1}{c|}{(0, 1/2, 1/2)} &  &
&  \\ \hline
\multirow{4}{*}{Ge2} & \multirow{4}{*}{2d} & \multicolumn{1}{c|}{%
\multirow{4}{*}{(1/3, 2/3, 1/2)}} & Ge3 & 4d & \multicolumn{1}{c|}{(1/3,
2/3, z)} & Ge3 & 4d & \multicolumn{1}{c|}{(1/3, 2/3, z)} & Ge5 & 2d &
\multicolumn{1}{c|}{(1/3, 2/3, z)} & Ge3 & 2d & (1/3, 2/3, z) \\
&  & \multicolumn{1}{c|}{} & Ge4 & 12g & \multicolumn{1}{c|}{(x, y, z)} & Ge4
& 12g & \multicolumn{1}{c|}{(x, y, z)} & Ge6 & 2d & \multicolumn{1}{c|}{
(1/3, 2/3, z)} & Ge4 & 2d & (1/3, 2/3, z) \\
&  & \multicolumn{1}{c|}{} &  &  & \multicolumn{1}{c|}{} &  &  &
\multicolumn{1}{c|}{} & Ge7 & 6i & \multicolumn{1}{c|}{(x, 2x, z)} & Ge5 & 6i
& (x, 2x, z) \\
&  & \multicolumn{1}{c|}{} &  &  & \multicolumn{1}{c|}{} &  &  &
\multicolumn{1}{c|}{} & Ge8 & 6i & \multicolumn{1}{c|}{(x, 2x, z)} & Ge6 & 6i
& (x, 2x, z) \\ \hline
\multirow{3}{*}{Fe} & \multirow{3}{*}{3f} & \multicolumn{1}{c|}{%
\multirow{3}{*}{(1/2, 0, 0)}} & Fe1 & 12g & \multicolumn{1}{c|}{(x, y, z)} &
Fe1 & 6f & \multicolumn{1}{c|}{(x, 0, 1/4)} & Fe1 & 6i & \multicolumn{1}{c|}{
(x, 2x, z)} & Fe1 & 6i & (x, 2x, z) \\
&  & \multicolumn{1}{c|}{} & Fe2 & 12g & \multicolumn{1}{c|}{(x, y, z)} & Fe2
& 6f & \multicolumn{1}{c|}{(x, 0, 1/4)} & Fe2 & 6i & \multicolumn{1}{c|}{(x,
2x, z)} & Fe2 & 6i & (x, 2x, z) \\
&  & \multicolumn{1}{c|}{} &  &  & \multicolumn{1}{c|}{} & Fe3 & 12g &
\multicolumn{1}{c|}{(x, y, z)} & Fe3 & 12j & \multicolumn{1}{c|}{(x, y, z)}
& Fe3 & 12j & (x, y, z) \\ \hline
&  &  &  &  &  &  &  &  &  &  &  &  &  &
\end{tabular}%
\end{table*}

\begin{table*}[htbp]
\caption{The corresponding Wyckoff positions and the coordinates of the
atoms in the pristine phase and CDW phases with different symmetries. (PART
\uppercase\expandafter{\romannumeral3}).}
\label{cdw3}%
\begin{tabular}{ccccccccccccccc}
\hline
\multicolumn{3}{c|}{Pristine phase(P6/mmm)} & \multicolumn{3}{c|}{SG163-P$%
\overline{3}$1c(type \uppercase\expandafter{\romannumeral1})} &
\multicolumn{3}{c|}{SG163-P$\overline{3}$1c(type \uppercase%
\expandafter{\romannumeral2})} & \multicolumn{3}{c|}{SG162-P$\overline{3}$%
1m(type \uppercase\expandafter{\romannumeral1})} & \multicolumn{3}{c}{SG162-P%
$\overline{3}$1m(type \uppercase\expandafter{\romannumeral2})} \\ \hline
& WP & \multicolumn{1}{c|}{Coordinates} &  & WP & \multicolumn{1}{c|}{
Coordinates} &  & WP & \multicolumn{1}{c|}{Coordinates} &  & WP &
\multicolumn{1}{c|}{Coordinates} &  & WP & Coordinates \\ \hline
\multirow{4}{*}{Ge1} & \multirow{4}{*}{1a} & \multicolumn{1}{c|}{%
\multirow{4}{*}{(0, 0, 0)}} & Ge1 & 2b & \multicolumn{1}{c|}{(0, 0, 0)} & Ge1
& 2a & \multicolumn{1}{c|}{(0, 0, 1/4)} & Ge1 & 1a & \multicolumn{1}{c|}{(0,
0, 0)} & Ge1 & 2e & (0, 0, z) \\
&  & \multicolumn{1}{c|}{} & Ge2 & 6g & \multicolumn{1}{c|}{(0, 1/2, 0)} &
Ge2 & 6h & \multicolumn{1}{c|}{(x, 2x, 1/4)} & Ge2 & 1b &
\multicolumn{1}{c|}{(0, 0, 1/2)} & Ge2 & 6k & (x, 0, z) \\
&  & \multicolumn{1}{c|}{} &  &  & \multicolumn{1}{c|}{} &  &  &
\multicolumn{1}{c|}{} & Ge3 & 3f & \multicolumn{1}{c|}{(1/2, 0, 0)} &  &  &
\\
&  & \multicolumn{1}{c|}{} &  &  & \multicolumn{1}{c|}{} &  &  &
\multicolumn{1}{c|}{} & Ge4 & 3g & \multicolumn{1}{c|}{(1/2, 0, 1/2)} &  &
&  \\ \hline
\multirow{4}{*}{Ge2} & \multirow{4}{*}{2d} & \multicolumn{1}{c|}{%
\multirow{4}{*}{(1/3, 2/3, 1/2)}} & Ge3 & 2c & \multicolumn{1}{c|}{(1/3,
2/3, 1/4)} & Ge3 & 4f & \multicolumn{1}{c|}{(1/3, 2/3, z)} & Ge5 & 4h &
\multicolumn{1}{c|}{(1/3, 2/3, z)} & Ge3 & 2c & (1/3, 2/3, 0) \\
&  & \multicolumn{1}{c|}{} & Ge4 & 2d & \multicolumn{1}{c|}{(1/3, 2/3, 3/4)}
& Ge4 & 12i & \multicolumn{1}{c|}{(x, y, z)} & Ge6 & 12l &
\multicolumn{1}{c|}{(x, y, z)} & Ge4 & 2d & (1/3, 2/3, 1/2) \\
&  & \multicolumn{1}{c|}{} & Ge5 & 6h & \multicolumn{1}{c|}{(x, 2x, 1/4)} &
&  & \multicolumn{1}{c|}{} &  &  & \multicolumn{1}{c|}{} & Ge5 & 6i & (x,
2x, 0) \\
&  & \multicolumn{1}{c|}{} & Ge6 & 6h & \multicolumn{1}{c|}{(x, 2x, 1/4)} &
&  & \multicolumn{1}{c|}{} &  &  & \multicolumn{1}{c|}{} & Ge6 & 6j & (x,
2x, 1/2) \\ \hline
\multirow{4}{*}{Fe} & \multirow{4}{*}{3f} & \multicolumn{1}{c|}{%
\multirow{4}{*}{(1/2, 0, 0))}} & Fe1 & 12i & \multicolumn{1}{c|}{(x, y, z)}
& Fe1 & 6h & \multicolumn{1}{c|}{(x, 2x, 1/4)} & Fe1 & 6i &
\multicolumn{1}{c|}{(x, 2x, 0)} & Fe1 & 6k & (x, 0, z) \\
&  & \multicolumn{1}{c|}{} & Fe2 & 12i & \multicolumn{1}{c|}{(x, y, z)} & Fe2
& 6h & \multicolumn{1}{c|}{(x, 2x, 1/4)} & Fe2 & 6j & \multicolumn{1}{c|}{
(x, 2x, 1/2)} & Fe2 & 6k & (x, 0, z) \\
&  & \multicolumn{1}{c|}{} &  &  & \multicolumn{1}{c|}{} & Fe3 & 12i &
\multicolumn{1}{c|}{(x, y, z)} & Fe3 & 6k & \multicolumn{1}{c|}{(x, 0, z)} &
Fe3 & 12i & (x, y, z) \\
&  & \multicolumn{1}{c|}{} &  &  & \multicolumn{1}{c|}{} &  &  &
\multicolumn{1}{c|}{} & Fe4 & 6k & \multicolumn{1}{c|}{(x, 0, z)} &  &  &
\\ \hline
&  &  &  &  &  &  &  &  &  &  &  &  &  &  \\ \hline
\multicolumn{3}{c|}{Pristine phase(P6/mmm)} & \multicolumn{3}{c|}{
SG68-Ccce(type \uppercase\expandafter{\romannumeral1})} &
\multicolumn{3}{c|}{SG68-Ccce(type \uppercase\expandafter{\romannumeral2})}
& \multicolumn{3}{c|}{SG68-Ccce(type \uppercase\expandafter{\romannumeral3})}
& \multicolumn{3}{c}{SG68-Ccce(type \uppercase\expandafter{\romannumeral4})}
\\ \hline
& WP & \multicolumn{1}{c|}{Coordinates} &  & WP & \multicolumn{1}{c|}{
Coordinates} &  & WP & \multicolumn{1}{c|}{Coordinates} &  & WP &
\multicolumn{1}{c|}{Coordinates} &  & WP & Coordinates \\ \hline
\multirow{3}{*}{Ge1} & \multirow{3}{*}{1a} & \multicolumn{1}{c|}{%
\multirow{3}{*}{(0, 0, 0)}} & Ge1 & 8c & \multicolumn{1}{c|}{(1/4, 1/4, 0)}
& Ge1 & 8e & \multicolumn{1}{c|}{(x, 1/4, 1/4)} & Ge1 & 8g &
\multicolumn{1}{c|}{(0, 1/4, z)} & Ge1 & 4a & (0, 1/4, 1/4) \\
&  & \multicolumn{1}{c|}{} & Ge2 & 8d & \multicolumn{1}{c|}{(0, 0, 0)} & Ge2
& 8f & \multicolumn{1}{c|}{(0, y, 1/4)} & Ge2 & 8h & \multicolumn{1}{c|}{
(1/4, 0, z)} & Ge2 & 4b & (0, 1/4, 3/4) \\
&  & \multicolumn{1}{c|}{} &  &  & \multicolumn{1}{c|}{} &  &  &
\multicolumn{1}{c|}{} &  &  & \multicolumn{1}{c|}{} & Ge3 & 8h & (1/4, 0, z)
\\ \hline
Ge2 & 2d & \multicolumn{1}{c|}{(1/3, 2/3, 1/2)} & Ge3 & 8f &
\multicolumn{1}{c|}{(0, y, 1/4)} & Ge3 & 16i & \multicolumn{1}{c|}{(x, y, z)}
& Ge3 & 8f & \multicolumn{1}{c|}{(0, y, 1/4)} & Ge3 & 16i & (x, y, z) \\
&  & \multicolumn{1}{c|}{} & Ge4 & 8f & \multicolumn{1}{c|}{(0, y, 1/4)} &
Ge4 & 16i & \multicolumn{1}{c|}{(x, y, z)} & Ge4 & 8f & \multicolumn{1}{c|}{
(0, y, 1/4)} & Ge4 & 16i & (x, y, z) \\
&  & \multicolumn{1}{c|}{} & Ge5 & 16i & \multicolumn{1}{c|}{(x, y, z)} &  &
& \multicolumn{1}{c|}{} & Ge5 & 16i & \multicolumn{1}{c|}{(x, y, z)} &  &  &
\\ \hline
\multirow{5}{*}{Fe} & \multirow{5}{*}{3f} & \multicolumn{1}{c|}{%
\multirow{5}{*}{(1/2, 0, 0))}} & Fe1 & 8g & \multicolumn{1}{c|}{(0, 1/4, z)}
& Fe1 & 4a & \multicolumn{1}{c|}{(0, 1/4, 1/4)} & Fe1 & 8c &
\multicolumn{1}{c|}{(1/4, 1/4, 0)} & Fe1 & 8e & (x, 1/4, 1/4) \\
&  & \multicolumn{1}{c|}{} & Fe2 & 8h & \multicolumn{1}{c|}{(1/4, 0, z)} &
Fe2 & 4b & \multicolumn{1}{c|}{(0, 1/4, 3/4)} & Fe2 & 8d &
\multicolumn{1}{c|}{(0, 0, 1/2)} & Fe2 & 8f & (0, y, 1/4) \\
&  & \multicolumn{1}{c|}{} & Fe3 & 16i & \multicolumn{1}{c|}{(x, y, z)} & Fe3
& 8h & \multicolumn{1}{c|}{(1/4, 0, z)} & Fe3 & 16i & \multicolumn{1}{c|}{
(x, y, z)} & Fe3 & 16i & (x, y, z) \\
&  & \multicolumn{1}{c|}{} & Fe4 & 16i & \multicolumn{1}{c|}{(x, y, z)} & Fe4
& 16i & \multicolumn{1}{c|}{(x, y, z)} & Fe4 & 16i & \multicolumn{1}{c|}{(x,
y, z)} & Fe4 & 16i & (x, y, z) \\
&  & \multicolumn{1}{c|}{} &  &  & \multicolumn{1}{c|}{} & Fe5 & 16i &
\multicolumn{1}{c|}{(x, y, z)} &  &  & \multicolumn{1}{c|}{} &  &  &  \\
\hline
&  &  &  &  &  &  &  &  &  &  &  &  &  &  \\ \hline
\multicolumn{3}{c|}{Pristine phase(P6/mmm)} & \multicolumn{3}{c|}{
SG67-Cmme(type \uppercase\expandafter{\romannumeral1})} &
\multicolumn{3}{c|}{SG67-Cmme(type \uppercase\expandafter{\romannumeral2})}
& \multicolumn{3}{c|}{SG67-Cmme(type \uppercase\expandafter{\romannumeral3})}
& \multicolumn{3}{c}{SG67-Cmme(type \uppercase\expandafter{\romannumeral4})}
\\ \hline
& WP & \multicolumn{1}{c|}{Coordinates} &  & WP & \multicolumn{1}{c|}{
Coordinates} &  & WP & \multicolumn{1}{c|}{Coordinates} &  & WP &
\multicolumn{1}{c|}{Coordinates} &  & WP & Coordinates \\ \hline
\multirow{4}{*}{Ge1} & \multirow{4}{*}{1a} & \multicolumn{1}{c|}{%
\multirow{4}{*}{(0, 0, 0)}} & Ge1 & 4c & \multicolumn{1}{c|}{(0, 0, 0)} & Ge1
& 4a & \multicolumn{1}{c|}{(1/4, 0, 0)} & Ge1 & 4g & \multicolumn{1}{c|}{(0,
1/4, z)} & Ge1 & 8n & (x, 1/4, z) \\
&  & \multicolumn{1}{c|}{} & Ge2 & 4d & \multicolumn{1}{c|}{(0, 0, 1/2)} &
Ge2 & 4b & \multicolumn{1}{c|}{(1/4, 0, 1/2)} & Ge2 & 4g &
\multicolumn{1}{c|}{(0, 1/4, z)} & Ge2 & 8m & (0, y, z) \\
&  & \multicolumn{1}{c|}{} & Ge3 & 4e & \multicolumn{1}{c|}{(1/4, 1/4, 0)} &
Ge3 & 4g & \multicolumn{1}{c|}{(0, 1/4, z)} & Ge3 & 8l & \multicolumn{1}{c|}{
(1/4, 0, z)} &  &  &  \\
&  & \multicolumn{1}{c|}{} & Ge4 & 4f & \multicolumn{1}{c|}{(1/4, 1/4, 1/2)}
& Ge4 & 4g & \multicolumn{1}{c|}{(0, 1/4, z)} &  &  & \multicolumn{1}{c|}{}
&  &  &  \\ \hline
\multirow{4}{*}{Ge2} & \multirow{4}{*}{2d} & \multicolumn{1}{c|}{%
\multirow{4}{*}{(1/3, 2/3, 1/2)}} & Ge5 & 8m & \multicolumn{1}{c|}{(0, y, z)}
& Ge5 & 8m & \multicolumn{1}{c|}{(0, y, z)} & Ge4 & 8j & \multicolumn{1}{c|}{
(1/4, y, 0)} & Ge3 & 8j & (1/4, y, 0) \\
&  & \multicolumn{1}{c|}{} & Ge6 & 8m & \multicolumn{1}{c|}{(0, y, z)} & Ge6
& 8m & \multicolumn{1}{c|}{(0, y, z)} & Ge5 & 8k & \multicolumn{1}{c|}{(1/4,
y, 1/2)} & Ge4 & 8k & (1/4, y, 1/2) \\
&  & \multicolumn{1}{c|}{} & Ge7 & 16o & \multicolumn{1}{c|}{(x, y, z)} & Ge7
& 16o & \multicolumn{1}{c|}{(x, y, z)} & Ge6 & 8m & \multicolumn{1}{c|}{(0,
y, z)} & Ge5 & 8m & (0, y, z) \\
&  & \multicolumn{1}{c|}{} &  &  & \multicolumn{1}{c|}{} &  &  &
\multicolumn{1}{c|}{} & Ge7 & 8m & \multicolumn{1}{c|}{(0, y, z)} & Ge6 & 8m
& (0, y, z) \\ \hline
\multirow{6}{*}{Fe} & \multirow{6}{*}{3f} & \multicolumn{1}{c|}{%
\multirow{6}{*}{(1/2, 0, 0))}} & Fe1 & 4a & \multicolumn{1}{c|}{(1/4, 0, 0)}
& Fe1 & 4c & \multicolumn{1}{c|}{(0, 0, 0)} & Fe1 & 8n & \multicolumn{1}{c|}{
(x, 1/4, z)} & Fe1 & 4g & (0, 1/4, z) \\
&  & \multicolumn{1}{c|}{} & Fe2 & 4b & \multicolumn{1}{c|}{(1/4, 0, 1/2)} &
Fe2 & 4d & \multicolumn{1}{c|}{(0, 0, 1/2)} & Fe2 & 8m & \multicolumn{1}{c|}{
(0, y, z)} & Fe2 & 4g & (0, 1/4, z) \\
&  & \multicolumn{1}{c|}{} & Fe3 & 4g & \multicolumn{1}{c|}{(0, 1/4, z)} &
Fe3 & 4e & \multicolumn{1}{c|}{(1/4, 1/4, 0)} & Fe3 & 16o &
\multicolumn{1}{c|}{(x, y, z)} & Fe3 & 8l & (1/4, 0, z) \\
&  & \multicolumn{1}{c|}{} & Fe4 & 4g & \multicolumn{1}{c|}{(0, 1/4, z)} &
Fe4 & 4f & \multicolumn{1}{c|}{(1/4, 1/4, 1/2)} & Fe4 & 16o &
\multicolumn{1}{c|}{(x, y, z)} & Fe4 & 16o & (x, y, z) \\
&  & \multicolumn{1}{c|}{} & Fe5 & 16o & \multicolumn{1}{c|}{(x, y, z)} & Fe5
& 16o & \multicolumn{1}{c|}{(x, y, z)} &  &  & \multicolumn{1}{c|}{} & Fe5 &
16o & (x, y, z) \\
&  & \multicolumn{1}{c|}{} & Fe6 & 16o & \multicolumn{1}{c|}{(x, y, z)} & Fe6
& 16o & \multicolumn{1}{c|}{(x, y, z)} &  &  & \multicolumn{1}{c|}{} &  &  &
\\ \hline
&  &  &  &  &  &  &  &  &  &  &  &  &  &  \\ \hline
\multicolumn{3}{c|}{Pristine phase(P6/mmm)} & \multicolumn{3}{c|}{
SG66-Cccm(type \uppercase\expandafter{\romannumeral1})} &
\multicolumn{3}{c|}{SG66-Cccm(type \uppercase\expandafter{\romannumeral2})}
& \multicolumn{3}{c|}{SG66-Cccm(type \uppercase\expandafter{\romannumeral3})}
& \multicolumn{3}{c}{SG66-Cccm(type \uppercase\expandafter{\romannumeral4})}
\\ \hline
& WP & \multicolumn{1}{c|}{Coordinates} &  & WP & \multicolumn{1}{c|}{
Coordinates} &  & WP & \multicolumn{1}{c|}{Coordinates} &  & WP &
\multicolumn{1}{c|}{Coordinates} &  & WP & Coordinates \\ \hline
\multirow{4}{*}{Ge1} & \multirow{4}{*}{1a} & \multicolumn{1}{c|}{%
\multirow{4}{*}{(0, 0, 0)}} & Ge1 & 4c & \multicolumn{1}{c|}{(0, 0, 0)} & Ge1
& 4a & \multicolumn{1}{c|}{(0, 0, 1/4)} & Ge1 & 8l & \multicolumn{1}{c|}{(x,
y, 0)} & Ge1 & 8g & (x, 0, 1/4) \\
&  & \multicolumn{1}{c|}{} & Ge2 & 4d & \multicolumn{1}{c|}{(0, 0, 1/2)} &
Ge2 & 4b & \multicolumn{1}{c|}{(0, 1/2, 1/4)} & Ge2 & 8l &
\multicolumn{1}{c|}{(x, y, 0)} & Ge2 & 8h & (0, y, 1/4) \\
&  & \multicolumn{1}{c|}{} & Ge3 & 4e & \multicolumn{1}{c|}{(1/4, 1/4, 0)} &
Ge3 & 8k & \multicolumn{1}{c|}{(1/4, 1/4, 1/4)} &  &  & \multicolumn{1}{c|}{}
&  &  &  \\
&  & \multicolumn{1}{c|}{} & Ge4 & 4f & \multicolumn{1}{c|}{(1/4, 1/4, 1/2)}
&  &  & \multicolumn{1}{c|}{} &  &  & \multicolumn{1}{c|}{} &  &  &  \\
\hline
\multirow{4}{*}{Ge2} & \multirow{4}{*}{2d} & \multicolumn{1}{c|}{%
\multirow{4}{*}{(1/3, 2/3, 1/2)}} & Ge5 & 8h & \multicolumn{1}{c|}{(0, y,
1/4)} & Ge4 & 8l & \multicolumn{1}{c|}{(x, y, 0)} & Ge3 & 8h &
\multicolumn{1}{c|}{(0, y, 1/4)} & Ge3 & 8l & (x, y, 0) \\
&  & \multicolumn{1}{c|}{} & Ge6 & 8h & \multicolumn{1}{c|}{(0, y, 1/4)} &
Ge5 & 8l & \multicolumn{1}{c|}{(x, y, 0)} & Ge4 & 8h & \multicolumn{1}{c|}{
(0, y, 1/4)} & Ge4 & 8l & (x, y, 0) \\
&  & \multicolumn{1}{c|}{} & Ge7 & 16m & \multicolumn{1}{c|}{(x, y, z)} & Ge6
& 8l & \multicolumn{1}{c|}{(x, y, 0)} & Ge5 & 16m & \multicolumn{1}{c|}{(x,
y, z)} & Ge5 & 8l & (x, y, 0) \\
&  & \multicolumn{1}{c|}{} &  &  & \multicolumn{1}{c|}{} & Ge7 & 8l &
\multicolumn{1}{c|}{(x, y, 0)} &  &  & \multicolumn{1}{c|}{} & Ge6 & 8l &
(x, y, 0) \\ \hline
\multirow{8}{*}{Fe} & \multirow{8}{*}{3f} & \multicolumn{1}{c|}{%
\multirow{8}{*}{(1/2, 0, 0))}} & Fe1 & 8l & \multicolumn{1}{c|}{(x, y, 0)} &
Fe1 & 8g & \multicolumn{1}{c|}{(x, 0, 1/4)} & Fe1 & 4c & \multicolumn{1}{c|}{
(0, 0, 0)} & Fe1 & 4a & (0, 0, 1/4) \\
&  & \multicolumn{1}{c|}{} & Fe2 & 8l & \multicolumn{1}{c|}{(x, y, 0)} & Fe2
& 8h & \multicolumn{1}{c|}{(0, y, 1/4)} & Fe2 & 4d & \multicolumn{1}{c|}{(0,
0, 1/2)} & Fe2 & 4b & (0, 1/2, 1/4) \\
&  & \multicolumn{1}{c|}{} & Fe3 & 8l & \multicolumn{1}{c|}{(x, y, 0)} & Fe3
& 16m & \multicolumn{1}{c|}{(x, y, z)} & Fe3 & 4e & \multicolumn{1}{c|}{
(1/4, 1/4, 0)} & Fe3 & 8k & (1/4, 1/4, 1/4) \\
&  & \multicolumn{1}{c|}{} & Fe4 & 8l & \multicolumn{1}{c|}{(x, y, 0)} & Fe4
& 16m & \multicolumn{1}{c|}{(x, y, z)} & Fe4 & 4f & \multicolumn{1}{c|}{
(1/4, 1/4, 1/2)} & Fe4 & 16m & (x, y, z) \\
&  & \multicolumn{1}{c|}{} & Fe5 & 8l & \multicolumn{1}{c|}{(x, y, 0)} &  &
& \multicolumn{1}{c|}{} & Fe5 & 8l & \multicolumn{1}{c|}{(x, y, 0)} & Fe5 &
16m & (x, y, z) \\
&  & \multicolumn{1}{c|}{} & Fe6 & 8l & \multicolumn{1}{c|}{(x, y, 0)} &  &
& \multicolumn{1}{c|}{} & Fe6 & 8l & \multicolumn{1}{c|}{(x, y, 0)} &  &  &
\\
&  & \multicolumn{1}{c|}{} &  &  & \multicolumn{1}{c|}{} &  &  &
\multicolumn{1}{c|}{} & Fe7 & 8l & \multicolumn{1}{c|}{(x, y, 0)} &  &  &
\\
&  & \multicolumn{1}{c|}{} &  &  & \multicolumn{1}{c|}{} &  &  &
\multicolumn{1}{c|}{} & Fe8 & 8l & \multicolumn{1}{c|}{(x, y, 0)} &  &  &
\\ \hline
\end{tabular}%
\end{table*}

\begin{table*}[htbp]
\caption{The corresponding Wyckoff positions and the coordinates of the
atoms in the pristine phase and CDW phases with different symmetries. (PART
\uppercase\expandafter{\romannumeral4}).}
\label{cdw4}%
\begin{tabular}{ccccccccccccccc}
\hline
\multicolumn{3}{c|}{Pristine phase(P6/mmm)} & \multicolumn{3}{c|}{
SG65-Cmmm(type \uppercase\expandafter{\romannumeral1})} &
\multicolumn{3}{c|}{SG65-Cmmm(case2)} & \multicolumn{3}{c|}{SG65-Cmmm(type %
\uppercase\expandafter{\romannumeral3})} & \multicolumn{3}{c}{SG65-Cmmm(type %
\uppercase\expandafter{\romannumeral4})} \\ \hline
& WP & \multicolumn{1}{c|}{Coordinates} &  & WP & \multicolumn{1}{c|}{
Coordinates} &  & WP & \multicolumn{1}{c|}{Coordinates} &  & WP &
\multicolumn{1}{c|}{Coordinates} &  & WP & Coordinates \\ \hline
\multirow{6}{*}{Ge1} & \multirow{6}{*}{1a} & \multicolumn{1}{c|}{%
\multirow{6}{*}{(0, 0, 0)}} & Ge1 & 2a & \multicolumn{1}{c|}{(0, 0, 0)} & Ge1
& 4i & \multicolumn{1}{c|}{(0, y, 0)} & Ge1 & 4k & \multicolumn{1}{c|}{(0,
0, z)} & Ge1 & 8o & (x, 0, z) \\
&  & \multicolumn{1}{c|}{} & Ge2 & 2b & \multicolumn{1}{c|}{(0, 1/2, 0)} &
Ge2 & 4j & \multicolumn{1}{c|}{(0, y, 1/2)} & Ge2 & 4l & \multicolumn{1}{c|}{
(0, 1/2, z)} & Ge2 & 8n & (0, y, z) \\
&  & \multicolumn{1}{c|}{} & Ge3 & 2c & \multicolumn{1}{c|}{(0, 1/2, 1/2)} &
Ge3 & 4g & \multicolumn{1}{c|}{(x, 0, 0)} & Ge3 & 8m & \multicolumn{1}{c|}{
(1/4, 1/4, z)} &  &  &  \\
&  & \multicolumn{1}{c|}{} & Ge4 & 2d & \multicolumn{1}{c|}{(0, 0, 1/2)} &
Ge4 & 4h & \multicolumn{1}{c|}{(x, 0, 1/2)} &  &  & \multicolumn{1}{c|}{} &
&  &  \\
&  & \multicolumn{1}{c|}{} & Ge5 & 4e & \multicolumn{1}{c|}{(1/4, 1/4, 0)} &
&  & \multicolumn{1}{c|}{} &  &  & \multicolumn{1}{c|}{} &  &  &  \\
&  & \multicolumn{1}{c|}{} & Ge6 & 4f & \multicolumn{1}{c|}{(1/4, 1/4, 1/2)}
&  &  & \multicolumn{1}{c|}{} &  &  & \multicolumn{1}{c|}{} &  &  &  \\
\hline
\multirow{6}{*}{Ge2} & \multirow{6}{*}{2d} & \multicolumn{1}{c|}{%
\multirow{6}{*}{(1/3, 2/3, 1/2)}} & Ge7 & 8n & \multicolumn{1}{c|}{(0, y, z)}
& Ge5 & 8n & \multicolumn{1}{c|}{(0, y, z)} & Ge4 & 4i & \multicolumn{1}{c|}{
(0, y, 0)} & Ge3 & 4i & (0, y, 0) \\
&  & \multicolumn{1}{c|}{} & Ge8 & 8n & \multicolumn{1}{c|}{(0, y, z)} & Ge6
& 8n & \multicolumn{1}{c|}{(0, y, z)} & Ge5 & 4i & \multicolumn{1}{c|}{(0,
y, 0)} & Ge4 & 4i & (0, y, 0) \\
&  & \multicolumn{1}{c|}{} & Ge9 & 16r & \multicolumn{1}{c|}{(x, y, z)} & Ge7
& 16r & \multicolumn{1}{c|}{(x, y, z)} & Ge6 & 4j & \multicolumn{1}{c|}{(0,
y, 1/2)} & Ge5 & 4j & (0, y, 1/2) \\
&  & \multicolumn{1}{c|}{} &  &  & \multicolumn{1}{c|}{} &  &  &
\multicolumn{1}{c|}{} & Ge7 & 4j & \multicolumn{1}{c|}{(0, y, 1/2)} & Ge6 &
4j & (0, y, 1/2) \\
&  & \multicolumn{1}{c|}{} &  &  & \multicolumn{1}{c|}{} &  &  &
\multicolumn{1}{c|}{} & Ge8 & 8p & \multicolumn{1}{c|}{(x, y, 0)} & Ge7 & 8p
& (x, y, 0) \\
&  & \multicolumn{1}{c|}{} &  &  & \multicolumn{1}{c|}{} &  &  &
\multicolumn{1}{c|}{} & Ge9 & 8q & \multicolumn{1}{c|}{(x, y, 1/2)} & Ge8 &
8q & (x, y, 1/2) \\ \hline
\multirow{10}{*}{Fe} & \multirow{10}{*}{3f} & \multicolumn{1}{c|}{%
\multirow{10}{*}{(1/2, 0, 0)}} & Fe1 & 4g & \multicolumn{1}{c|}{(x, 0, 0)} &
Fe1 & 2a & \multicolumn{1}{c|}{(0, 0, 0)} & Fe1 & 8o & \multicolumn{1}{c|}{
(x, 0, z)} & Fe1 & 4k & (0, 0, z) \\
&  & \multicolumn{1}{c|}{} & Fe2 & 4h & \multicolumn{1}{c|}{(x, 0, 1/2)} &
Fe2 & 2b & \multicolumn{1}{c|}{(0, 1/2, 0)} & Fe2 & 8n & \multicolumn{1}{c|}{
(0, y, z)} & Fe2 & 4l & (0, 1/2, z) \\
&  & \multicolumn{1}{c|}{} & Fe3 & 4i & \multicolumn{1}{c|}{(0, y, 0)} & Fe3
& 2c & \multicolumn{1}{c|}{(0, 1/2, 1/2)} & Fe3 & 16r & \multicolumn{1}{c|}{
(x, y, z)} & Fe3 & 8m & (1/4, 1/4, z) \\
&  & \multicolumn{1}{c|}{} & Fe4 & 4j & \multicolumn{1}{c|}{(0, y, 1/2)} &
Fe4 & 2d & \multicolumn{1}{c|}{(0, 0, 1/2)} & Fe4 & 16r &
\multicolumn{1}{c|}{(x, y, z)} & Fe4 & 16r & (x, y, z) \\
&  & \multicolumn{1}{c|}{} & Fe5 & 8p & \multicolumn{1}{c|}{(x, y, 0)} & Fe5
& 4e & \multicolumn{1}{c|}{(1/4, 1/4, 0)} &  &  & \multicolumn{1}{c|}{} & Fe5
& 16r & (x, y, z) \\
&  & \multicolumn{1}{c|}{} & Fe6 & 8p & \multicolumn{1}{c|}{(x, y, 0)} & Fe6
& 4f & \multicolumn{1}{c|}{(1/4, 1/4, 1/2)} &  &  & \multicolumn{1}{c|}{} &
&  &  \\
&  & \multicolumn{1}{c|}{} & Fe7 & 8q & \multicolumn{1}{c|}{(x, y, 1/2)} &
Fe7 & 8p & \multicolumn{1}{c|}{(x, y, 0)} &  &  & \multicolumn{1}{c|}{} &  &
&  \\
&  & \multicolumn{1}{c|}{} & Fe8 & 8q & \multicolumn{1}{c|}{(x, y, 1/2)} &
Fe8 & 8p & \multicolumn{1}{c|}{(x, y, 0)} &  &  & \multicolumn{1}{c|}{} &  &
&  \\
&  & \multicolumn{1}{c|}{} &  &  & \multicolumn{1}{c|}{} & Fe9 & 8q &
\multicolumn{1}{c|}{(x, y, 1/2)} &  &  & \multicolumn{1}{c|}{} &  &  &  \\
&  & \multicolumn{1}{c|}{} &  &  & \multicolumn{1}{c|}{} & Fe10 & 8q &
\multicolumn{1}{c|}{(x, y, 1/2)} &  &  & \multicolumn{1}{c|}{} &  &  &  \\
\hline
&  &  &  &  &  &  &  &  &  &  &  &  &  &  \\ \hline
\multicolumn{3}{c|}{Pristine phase(P6/mmm)} & \multicolumn{3}{c|}{
SG64-Cmce(type \uppercase\expandafter{\romannumeral1})} &
\multicolumn{3}{c|}{SG64-Cmce(type \uppercase\expandafter{\romannumeral2})}
& \multicolumn{3}{c|}{SG64-Cmce(type \uppercase\expandafter{\romannumeral3})}
& \multicolumn{3}{c}{SG64-Cmce(type \uppercase\expandafter{\romannumeral4})}
\\ \hline
& WP & \multicolumn{1}{c|}{Coordinates} &  & WP & \multicolumn{1}{c|}{
Coordinates} &  & WP & \multicolumn{1}{c|}{Coordinates} &  & WP &
\multicolumn{1}{c|}{Coordinates} &  & WP & Coordinates \\ \hline
\multirow{3}{*}{Ge1} & \multirow{3}{*}{1a} & \multicolumn{1}{c|}{%
\multirow{3}{*}{(0, 0, 0)}} & Ge1 & 4a & \multicolumn{1}{c|}{(0, 0, 0)} & Ge1
& 4a & \multicolumn{1}{c|}{(0, 0, 0)} & Ge1 & 8e & \multicolumn{1}{c|}{(1/4,
y, 1/4)} & Ge1 & 8e & (1/4, y, 1/4) \\
&  & \multicolumn{1}{c|}{} & Ge2 & 4b & \multicolumn{1}{c|}{(0, 0, 1/2)} &
Ge2 & 4b & \multicolumn{1}{c|}{(0, 0, 1/2)} & Ge2 & 8f & \multicolumn{1}{c|}{
(0, y, z)} & Ge2 & 8f & (0, y, z) \\
&  & \multicolumn{1}{c|}{} & Ge3 & 8c & \multicolumn{1}{c|}{(1/4, 1/4, 0)} &
Ge3 & 8c & \multicolumn{1}{c|}{(1/4, 1/4, 0)} &  &  & \multicolumn{1}{c|}{}
&  &  &  \\ \hline
\multirow{4}{*}{Ge2} & \multirow{4}{*}{2d} & \multicolumn{1}{c|}{%
\multirow{4}{*}{(1/3, 2/3, 1/2)}} & Ge4 & 16g & \multicolumn{1}{c|}{(x, y, z)
} & Ge4 & 8e & \multicolumn{1}{c|}{(1/4, y, 1/4)} & Ge3 & 8d &
\multicolumn{1}{c|}{(x, 0, 0)} & Ge3 & 8f & (0, y, z) \\
&  & \multicolumn{1}{c|}{} & Ge5 & 16g & \multicolumn{1}{c|}{(x, y, z)} & Ge5
& 8e & \multicolumn{1}{c|}{(1/4, y, 1/4)} & Ge4 & 8d & \multicolumn{1}{c|}{
(x, 0, 0)} & Ge4 & 8f & (0, y, z) \\
&  & \multicolumn{1}{c|}{} &  &  & \multicolumn{1}{c|}{} & Ge6 & 8f &
\multicolumn{1}{c|}{(0, y, z)} & Ge5 & 16g & \multicolumn{1}{c|}{(x, y, z)}
& Ge5 & 16g & (x, y, z) \\
&  & \multicolumn{1}{c|}{} &  &  & \multicolumn{1}{c|}{} & Ge7 & 8f &
\multicolumn{1}{c|}{(0, y, z)} &  &  & \multicolumn{1}{c|}{} &  &  &  \\
\hline
\multirow{5}{*}{Fe} & \multirow{5}{*}{3f} & \multicolumn{1}{c|}{%
\multirow{5}{*}{(1/2, 0, 0))}} & Fe1 & 8d & \multicolumn{1}{c|}{(x, 0, 0)} &
Fe1 & 8e & \multicolumn{1}{c|}{(1/4, y, 1/4)} & Fe1 & 8e &
\multicolumn{1}{c|}{(1/4, y, 1/4)} & Fe1 & 8e & (1/4, y, 1/4) \\
&  & \multicolumn{1}{c|}{} & Fe2 & 8f & \multicolumn{1}{c|}{(0, y, z)} & Fe2
& 8f & \multicolumn{1}{c|}{(0, y, z)} & Fe2 & 8f & \multicolumn{1}{c|}{(0,
y, z)} & Fe2 & 8f & (0, y, z) \\
&  & \multicolumn{1}{c|}{} & Fe3 & 16g & \multicolumn{1}{c|}{(x, y, z)} & Fe3
& 16g & \multicolumn{1}{c|}{(x, y, z)} & Fe3 & 16g & \multicolumn{1}{c|}{(x,
y, z)} & Fe3 & 16g & (x, y, z) \\
&  & \multicolumn{1}{c|}{} & Fe4 & 16g & \multicolumn{1}{c|}{(x, y, z)} & Fe4
& 16g & \multicolumn{1}{c|}{(x, y, z)} & Fe4 & 16g & \multicolumn{1}{c|}{(x,
y, z)} & Fe4 & 16g & (x, y, z) \\
&  & \multicolumn{1}{c|}{} &  &  & \multicolumn{1}{c|}{} &  &  &
\multicolumn{1}{c|}{} &  &  & \multicolumn{1}{c|}{} &  &  &  \\ \hline
&  &  &  &  &  &  &  &  &  &  &  &  &  &  \\ \hline
\multicolumn{3}{c|}{Pristine phase(P6/mmm)} & \multicolumn{3}{c|}{
SG64-Cmce(type \uppercase\expandafter{\romannumeral5})} &
\multicolumn{3}{c|}{SG64-Cmce(type \uppercase\expandafter{\romannumeral6})}
& \multicolumn{3}{c|}{SG64-Cmce(type \uppercase\expandafter{\romannumeral7})}
& \multicolumn{3}{c}{SG64-Cmce(type \uppercase\expandafter{\romannumeral8})}
\\ \hline
& WP & \multicolumn{1}{c|}{Coordinates} &  & WP & \multicolumn{1}{c|}{
Coordinates} &  & WP & \multicolumn{1}{c|}{Coordinates} &  & WP &
\multicolumn{1}{c|}{Coordinates} &  & WP & Coordinates \\ \hline
\multirow{3}{*}{Ge1} & \multirow{3}{*}{1a} & \multicolumn{1}{c|}{%
\multirow{3}{*}{(0, 0, 0)}} & Ge1 & 8e & \multicolumn{1}{c|}{(1/4, y, 1/4)}
& Ge1 & 8e & \multicolumn{1}{c|}{(1/4, y, 1/4)} & Ge1 & 8d &
\multicolumn{1}{c|}{(x, 0, 0)} & Ge1 & 8d & (x, 0, 0) \\
&  & \multicolumn{1}{c|}{} & Ge2 & 8f & \multicolumn{1}{c|}{(0, y, z)} & Ge2
& 8f & \multicolumn{1}{c|}{(0, y, z)} & Ge2 & 8f & \multicolumn{1}{c|}{(0,
y, z)} & Ge2 & 8f & (0, y, z) \\
&  & \multicolumn{1}{c|}{} &  &  & \multicolumn{1}{c|}{} &  &  &
\multicolumn{1}{c|}{} &  &  & \multicolumn{1}{c|}{} &  &  &  \\ \hline
\multirow{4}{*}{Ge2} & \multirow{4}{*}{2d} & \multicolumn{1}{c|}{%
\multirow{4}{*}{(1/3, 2/3, 1/2)}} & Ge3 & 8f & \multicolumn{1}{c|}{(0, y, z)}
& Ge3 & 8d & \multicolumn{1}{c|}{(x, 0, 0)} & Ge3 & 16g &
\multicolumn{1}{c|}{(x, y, z)} & Ge4 & 8e & (1/4, y, 1/4) \\
&  & \multicolumn{1}{c|}{} & Ge4 & 8f & \multicolumn{1}{c|}{(0, y, z)} & Ge4
& 8d & \multicolumn{1}{c|}{(x, 0, 0)} & Ge4 & 16g & \multicolumn{1}{c|}{(x,
y, z)} & Ge5 & 8e & (1/4, y, 1/4) \\
&  & \multicolumn{1}{c|}{} & Ge5 & 16g & \multicolumn{1}{c|}{(x, y, z)} & Ge5
& 16g & \multicolumn{1}{c|}{(x, y, z)} &  &  & \multicolumn{1}{c|}{} & Ge6 &
8f & (0, y, z) \\
&  & \multicolumn{1}{c|}{} &  &  & \multicolumn{1}{c|}{} &  &  &
\multicolumn{1}{c|}{} &  &  & \multicolumn{1}{c|}{} & Ge7 & 8f & (0, y, z)
\\ \hline
Fe & 3f & \multicolumn{1}{c|}{(1/2, 0, 0))} & Fe1 & 8e & \multicolumn{1}{c|}{
(1/4, y, 1/4)} & Fe1 & 8e & \multicolumn{1}{c|}{(1/4, y, 1/4)} & Fe1 & 4a &
\multicolumn{1}{c|}{(0, 0, 0)} & Fe1 & 4a & (0, 0, 0) \\
\multirow{4}{*}{} & \multirow{4}{*}{} & \multicolumn{1}{c|}{\multirow{4}{*}{}
} & Fe2 & 8f & \multicolumn{1}{c|}{(0, y, z)} & Fe2 & 8f &
\multicolumn{1}{c|}{(0, y, z)} & Fe2 & 4b & \multicolumn{1}{c|}{(0, 0, 1/2)}
& Fe2 & 4b & (0, 0, 1/2) \\
&  & \multicolumn{1}{c|}{} & Fe3 & 16g & \multicolumn{1}{c|}{(x, y, z)} & Fe3
& 16g & \multicolumn{1}{c|}{(x, y, z)} & Fe3 & 8c & \multicolumn{1}{c|}{
(1/4, 1/4, 0)} & Fe3 & 8c & (1/4, 1/4, 0) \\
&  & \multicolumn{1}{c|}{} & Fe4 & 16g & \multicolumn{1}{c|}{(x, y, z)} & Fe4
& 16g & \multicolumn{1}{c|}{(x, y, z)} & Fe4 & 16g & \multicolumn{1}{c|}{(x,
y, z)} & Fe4 & 16g & (x, y, z) \\
&  & \multicolumn{1}{c|}{} &  &  & \multicolumn{1}{c|}{} &  &  &
\multicolumn{1}{c|}{} & Fe5 & 16g & \multicolumn{1}{c|}{(x, y, z)} & Fe5 &
16g & (x, y, z) \\ \hline
\end{tabular}%
\end{table*}

\begin{table*}[htbp]
\caption{The corresponding Wyckoff positions and the coordinates of the
atoms in the pristine phase and CDW phases with different symmetries. (PART
\uppercase\expandafter{\romannumeral5}).}
\label{cdw5}%
\begin{tabular}{ccccccccccccccc}
\hline
\multicolumn{3}{c|}{Pristine phase(P6/mmm)} & \multicolumn{3}{c|}{
SG63-Cmcm(type\uppercase\expandafter{\romannumeral1})} & \multicolumn{3}{c|}{
SG63-Cmcm(type \uppercase\expandafter{\romannumeral2})} &
\multicolumn{3}{c|}{SG63-Cmcm(type \uppercase\expandafter{\romannumeral3})}
& \multicolumn{3}{c}{SG63-Cmcm(type \uppercase\expandafter{\romannumeral4})}
\\ \hline
& WP & \multicolumn{1}{c|}{Coordinates} &  & WP & \multicolumn{1}{c|}{
Coordinates} &  & WP & \multicolumn{1}{c|}{Coordinates} &  & WP &
\multicolumn{1}{c|}{Coordinates} &  & WP & Coordinates \\ \hline
\multirow{3}{*}{Ge1} & \multirow{3}{*}{1a} & \multicolumn{1}{c|}{%
\multirow{3}{*}{(0, 0, 0)}} & Ge1 & 4a & \multicolumn{1}{c|}{(0, 0, 0)} & Ge1
& 4a & \multicolumn{1}{c|}{(0, 0, 0)} & Ge1 & 4c & \multicolumn{1}{c|}{(0,
y, 1/4)} & Ge1 & 4c & (0, y, 1/4) \\
&  & \multicolumn{1}{c|}{} & Ge2 & 4b & \multicolumn{1}{c|}{(0, 1/2, 0)} &
Ge2 & 4b & \multicolumn{1}{c|}{(0, 1/2, 0)} & Ge2 & 4c & \multicolumn{1}{c|}{
(0, y, 1/4)} & Ge2 & 4c & (0, y, 1/4) \\
&  & \multicolumn{1}{c|}{} & Ge3 & 8d & \multicolumn{1}{c|}{(1/4, 1/4, 0)} &
Ge3 & 8d & \multicolumn{1}{c|}{(1/4, 1/4, 0)} & Ge3 & 8g &
\multicolumn{1}{c|}{(x, y, 1/4)} & Ge3 & 8g & (x, y, 1/4) \\ \hline
\multirow{6}{*}{Ge2} & \multirow{6}{*}{2d} & \multicolumn{1}{c|}{%
\multirow{6}{*}{(1/3, 2/3, 1/2)}} & Ge4 & 8g & \multicolumn{1}{c|}{(x, y,
1/4)} & Ge4 & 4c & \multicolumn{1}{c|}{(0, y, 1/4)} & Ge4 & 8e &
\multicolumn{1}{c|}{(x, 0, 0)} & Ge4 & 8e & (x, 0, 0) \\
&  & \multicolumn{1}{c|}{} & Ge5 & 8g & \multicolumn{1}{c|}{(x, y, 1/4)} &
Ge5 & 4c & \multicolumn{1}{c|}{(0, y, 1/4)} & Ge5 & 8e & \multicolumn{1}{c|}{
(x, 0, 0)} & Ge5 & 8e & (x, 0, 0) \\
&  & \multicolumn{1}{c|}{} & Ge6 & 8g & \multicolumn{1}{c|}{(x, y, 1/4)} &
Ge6 & 4c & \multicolumn{1}{c|}{(0, y, 1/4)} & Ge6 & 16h &
\multicolumn{1}{c|}{(x, y, z)} & Ge6 & 16h & (x, y, z) \\
&  & \multicolumn{1}{c|}{} & Ge7 & 8g & \multicolumn{1}{c|}{(x, y, 1/4)} &
Ge7 & 4c & \multicolumn{1}{c|}{(0, y, 1/4)} & \multicolumn{1}{l}{} &
\multicolumn{1}{l}{} & \multicolumn{1}{l|}{} & \multicolumn{1}{l}{} &
\multicolumn{1}{l}{} & \multicolumn{1}{l}{} \\
&  & \multicolumn{1}{c|}{} & \multicolumn{1}{l}{} & \multicolumn{1}{l}{} &
\multicolumn{1}{l|}{} & Ge8 & 8g & \multicolumn{1}{c|}{(x, y, 1/4)} &
\multicolumn{1}{l}{} & \multicolumn{1}{l}{} & \multicolumn{1}{l|}{} &
\multicolumn{1}{l}{} & \multicolumn{1}{l}{} & \multicolumn{1}{l}{} \\
&  & \multicolumn{1}{c|}{} &  &  & \multicolumn{1}{c|}{} & Ge9 & 8g &
\multicolumn{1}{c|}{(x, y, 1/4)} &  &  & \multicolumn{1}{c|}{} &  &  &  \\
\hline
\multirow{7}{*}{Fe} & \multirow{7}{*}{3f} & \multicolumn{1}{c|}{%
\multirow{7}{*}{(1/2, 0, 0))}} & Fe1 & 8e & \multicolumn{1}{c|}{(x, 0, 0)} &
Fe1 & 8e & \multicolumn{1}{c|}{(x, 0, 0)} & Fe1 & 4c & \multicolumn{1}{c|}{
(0, y, 1/4)} & Fe1 & 4c & (0, y, 1/4) \\
&  & \multicolumn{1}{c|}{} & Fe2 & 8f & \multicolumn{1}{c|}{(0, y, z)} & Fe2
& 8f & \multicolumn{1}{c|}{(0, y, z)} & Fe2 & 4c & \multicolumn{1}{c|}{(0,
y, 1/4)} & Fe2 & 4c & (0, y, 1/4) \\
&  & \multicolumn{1}{c|}{} & Fe3 & 16h & \multicolumn{1}{c|}{(x, y, z)} & Fe3
& 16h & \multicolumn{1}{c|}{(x, y, z)} & Fe3 & 8g & \multicolumn{1}{c|}{(x,
y, 1/4)} & Fe3 & 8g & (x, y, 1/4) \\
&  & \multicolumn{1}{c|}{} & Fe4 & 16h & \multicolumn{1}{c|}{(x, y, z)} & Fe4
& 16h & \multicolumn{1}{c|}{(x, y, z)} & Fe4 & 8g & \multicolumn{1}{c|}{(x,
y, 1/4)} & Fe4 & 8g & (x, y, 1/4) \\
&  & \multicolumn{1}{c|}{} &  &  & \multicolumn{1}{c|}{} &  &  &
\multicolumn{1}{c|}{} & Fe5 & 8g & \multicolumn{1}{c|}{(x, y, 1/4)} & Fe5 &
8g & (x, y, 1/4) \\
&  & \multicolumn{1}{c|}{} &  &  & \multicolumn{1}{c|}{} &  &  &
\multicolumn{1}{c|}{} & Fe6 & 8g & \multicolumn{1}{c|}{(x, y, 1/4)} & Fe6 &
8g & (x, y, 1/4) \\
&  & \multicolumn{1}{c|}{} &  &  & \multicolumn{1}{c|}{} &  &  &
\multicolumn{1}{c|}{} & Fe7 & 8g & \multicolumn{1}{c|}{(x, y, 1/4)} & Fe7 &
8g & (x, y, 1/4) \\ \hline
&  &  &  &  &  &  &  &  &  &  &  &  &  &  \\ \hline
\multicolumn{3}{c|}{Pristine phase(P6/mmm)} & \multicolumn{3}{c|}{
SG63-Cmcm(type \uppercase\expandafter{\romannumeral5})} &
\multicolumn{3}{c|}{SG63-Cmcm(type \uppercase\expandafter{\romannumeral6})}
& \multicolumn{3}{c|}{SG63-Cmcm(type \uppercase\expandafter{\romannumeral7})}
& \multicolumn{3}{c}{SG63-Cmcm(type \uppercase\expandafter{\romannumeral8})}
\\ \hline
& WP & \multicolumn{1}{c|}{Coordinates} &  & WP & \multicolumn{1}{c|}{
Coordinates} &  & WP & \multicolumn{1}{c|}{Coordinates} &  & WP &
\multicolumn{1}{c|}{Coordinates} &  & WP & Coordinates \\ \hline
\multirow{3}{*}{Ge1} & \multirow{3}{*}{1a} & \multicolumn{1}{c|}{%
\multirow{3}{*}{(0, 0, 0)}} & Ge1 & 4c & \multicolumn{1}{c|}{(0, y, 1/4)} &
Ge1 & 4c & \multicolumn{1}{c|}{(0, y, 1/4)} & Ge1 & 8e & \multicolumn{1}{c|}{
(x, 0, 0)} & Ge1 & 8e & (x, 0, 0) \\
&  & \multicolumn{1}{c|}{} & Ge2 & 4c & \multicolumn{1}{c|}{(0, y, 1/4)} &
Ge2 & 4c & \multicolumn{1}{c|}{(0, y, 1/4)} & Ge2 & 8f & \multicolumn{1}{c|}{
(0, y, z)} & Ge2 & 8f & (0, y, z) \\
&  & \multicolumn{1}{c|}{} & Ge3 & 8g & \multicolumn{1}{c|}{(x, y, 1/4)} &
Ge3 & 8g & \multicolumn{1}{c|}{(x, y, 1/4)} &  &  & \multicolumn{1}{c|}{} &
&  &  \\ \hline
\multirow{6}{*}{Ge2} & \multirow{6}{*}{2d} & \multicolumn{1}{c|}{%
\multirow{6}{*}{(1/3, 2/3, 1/2)}} & Ge4 & 8f & \multicolumn{1}{c|}{(0, y, z)}
& Ge4 & 8f & \multicolumn{1}{c|}{(0, y, z)} & Ge3 & 8g & \multicolumn{1}{c|}{
(x, y, 1/4)} & Ge4 & 4c & (0, y, 1/4) \\
&  & \multicolumn{1}{c|}{} & Ge5 & 8f & \multicolumn{1}{c|}{(0, y, z)} & Ge5
& 8f & \multicolumn{1}{c|}{(0, y, z)} & Ge4 & 8g & \multicolumn{1}{c|}{(x,
y, 1/4)} & Ge5 & 4c & (0, y, 1/4) \\
&  & \multicolumn{1}{c|}{} & Ge6 & 16h & \multicolumn{1}{c|}{(x, y, z)} & Ge6
& 16h & \multicolumn{1}{c|}{(x, y, z)} & Ge5 & 8g & \multicolumn{1}{c|}{(x,
y, 1/4)} & Ge6 & 4c & (0, y, 1/4) \\
&  & \multicolumn{1}{c|}{} &  &  & \multicolumn{1}{c|}{} &  &  &
\multicolumn{1}{c|}{} & Ge6 & 8g & \multicolumn{1}{c|}{(x, y, 1/4)} & Ge7 &
4c & (0, y, 1/4) \\
&  & \multicolumn{1}{c|}{} &  &  & \multicolumn{1}{c|}{} &  &  &
\multicolumn{1}{c|}{} &  &  & \multicolumn{1}{c|}{} & Ge8 & 8g & (x, y, 1/4)
\\
&  & \multicolumn{1}{c|}{} &  &  & \multicolumn{1}{c|}{} &  &  &
\multicolumn{1}{c|}{} &  &  & \multicolumn{1}{c|}{} & Ge9 & 8g & (x, y, 1/4)
\\ \hline
\multirow{7}{*}{Fe} & \multirow{7}{*}{3f} & \multicolumn{1}{c|}{%
\multirow{7}{*}{(1/2, 0, 0))}} & Fe1 & 4c & \multicolumn{1}{c|}{(0, y, 1/4)}
& Fe1 & 4c & \multicolumn{1}{c|}{(0, y, 1/4)} & Fe1 & 4a &
\multicolumn{1}{c|}{(0, 0, 0)} & Fe1 & 4a & (0, 0, 0) \\
&  & \multicolumn{1}{c|}{} & Fe2 & 4c & \multicolumn{1}{c|}{(0, y, 1/4)} &
Fe2 & 4c & \multicolumn{1}{c|}{(0, y, 1/4)} & Fe2 & 4b & \multicolumn{1}{c|}{
(0, 1/2, 0)} & Fe2 & 4b & (0, 1/2, 0) \\
&  & \multicolumn{1}{c|}{} & Fe3 & 8g & \multicolumn{1}{c|}{(x, y, 1/4)} &
Fe3 & 8g & \multicolumn{1}{c|}{(x, y, 1/4)} & Fe3 & 8d & \multicolumn{1}{c|}{
(1/4, 1/4, 0)} & Fe3 & 8d & (1/4, 1/4, 0) \\
&  & \multicolumn{1}{c|}{} & Fe4 & 8g & \multicolumn{1}{c|}{(x, y, 1/4)} &
Fe4 & 8g & \multicolumn{1}{c|}{(x, y, 1/4)} & Fe4 & 16h &
\multicolumn{1}{c|}{(x, y, z)} & Fe4 & 16h & (x, y, z) \\
&  & \multicolumn{1}{c|}{} & Fe5 & 8g & \multicolumn{1}{c|}{(x, y, 1/4)} &
Fe5 & 8g & \multicolumn{1}{c|}{(x, y, 1/4)} & Fe5 & 16h &
\multicolumn{1}{c|}{(x, y, z)} & Fe5 & 16h & (x, y, z) \\
&  & \multicolumn{1}{c|}{} & Fe6 & 8g & \multicolumn{1}{c|}{(x, y, 1/4)} &
Fe6 & 8g & \multicolumn{1}{c|}{(x, y, 1/4)} &  &  & \multicolumn{1}{c|}{} &
&  &  \\
&  & \multicolumn{1}{c|}{} & Fe7 & 8g & \multicolumn{1}{c|}{(x, y, 1/4)} &
Fe7 & 8g & \multicolumn{1}{c|}{(x, y, 1/4)} &  &  & \multicolumn{1}{c|}{} &
&  &  \\ \hline
\end{tabular}%
\end{table*}

\clearpage

\bibliographystyle{aps}
\bibliography{FeGe}

\end{document}